\documentclass[useAMS,usenatbib]{mn2e}

\usepackage[T1]{fontenc}
 \usepackage{pslatex}
\usepackage{amsmath}
\usepackage{amsfonts}
\usepackage{color}
\usepackage{url}
\usepackage{graphicx}
\usepackage{subfigure}
\usepackage{amssymb}
\usepackage{soul}
{\newif\ifnotend
\notendtrue
\def\veclist{ABCDEFGHIJKLMNOPQRSTUVWXYZabcdefghijklmnopqrstuvwxyz.}
\def\top#1#2.{#1}
\def\tail#1#2.{#2.}
\loop\expandafter\xdef\csname v\expandafter\top\veclist\endcsname%
{{\noexpand\bf\expandafter\top\veclist}}
\edef\veclist{\expandafter\tail\veclist}
\if\veclist.\notendfalse\fi\ifnotend\repeat}
\def\pa{\partial}

\mathchardef\mhyphen="2D

\title[Gas flow in barred potentials II]{Gas flow in barred potentials II. Bar Driven Spiral Arms.}
\author[Sormani, Binney \& Magorrian]{Mattia C. Sormani$^1$, James Binney$^1$ and John Magorrian$^{1,2}$\\
$^1$ Rudolf Peierls Centre for Theoretical Physics, 1 Keble Road, Oxford
OX1 3NP\\
$^2$ Institut d'Astrophysique de Paris, 98bis boulevard Arago, 75014 Paris}
\begin{document}

\date{}

\def\p{\partial}
\def\Omegap{\Omega_{\rm p}}

\newcommand{\di}{\mathrm{d}}
\newcommand{\bfx}{\mathbf{x}}
\newcommand{\bfe}{\mathbf{e}}
\newcommand{\vlos}{\mathrm{v}_{\rm los}}
\newcommand{\Tspin}{T_{\rm s}}
\newcommand{\Tb}{T_{\rm b}}
\newcommand{\degree}{\ensuremath{^\circ}}
\newcommand{\Th}{T_{\rm h}}
\newcommand{\Tc}{T_{\rm c}}
\newcommand{\bfr}{\mathbf{r}}
\newcommand{\bfv}{\mathbf{v}}
\newcommand{\pc}{\,{\rm pc}}
\newcommand{\kpc}{\,{\rm kpc}}
\newcommand{\Myr}{\,{\rm Myr}}
\newcommand{\Gyr}{\,{\rm Gyr}}
\newcommand{\kms}{\,{\rm km\, s^{-1}}}
\newcommand{\de}[2]{\frac{\partial #1}{\partial {#2}}}
\newcommand{\cs}{c_{\rm s}}

\maketitle

\begin{abstract}
Spiral arms that emerge from the ends of a galactic bar are important in
interpreting observations of our and external galaxies. It is therefore
important to understand the physical mechanism that causes them. We find that
these spiral arms can be understood as kinematic density waves generated by
librations around underlying ballistic closed orbits. This is even true in
the case of a strong bar, provided the librations are around the appropriate
closed orbits and not around the circular orbits that form the basis of the
epicycle approximation. An important consequence is that it is a potential's  orbital
structure that determines whether a bar
should be classified as weak or strong, and not crude estimates of the
potential's deviation from axisymmetry.
\end{abstract}


\section{Introduction} \label{sec:introduction}

Many grand design barred spirals exhibit spiral arms starting at and
extending out from the end of the bar. How are these spiral arms generated?
The natural interpretation is that they are driven by the bar. The spiral
arms and the bar must also be rotating at the same pattern speed if the
starting points of the arms always coincide with the bar's ends.  Such spiral
arms are relevant for the interpretation of observations of both our Galaxy
and external galaxies. For example, it is likely that some features present
in spectral-line emission of atomic and molecular gas in the inner Galaxy,
for instance the ``3kpc arm'', are produced by such arms
\citep[e.g.]{Dame2008}. It is therefore
important to have a physical understanding of why such spiral arms arise.

On the theoretical side, the possibility of bar-driven spiral arms is now
well established. The first investigations used analytical methods and were
carried out by \cite{FeldmanLin1973} and \cite{LinLau1975}. Then, in the
following years, \cite{SandersHuntley1976,Sanders1977} and
\cite{Huntley++1978} performed pioneering numerical experiments and discussed
the physical mechanism responsible for the generation of the spiral arms in
terms of closed orbits. Their work was later extended independently by
\cite{Wada1994} and \cite{LindbladLindblad1994}, and more recently by
\cite{PinolFerrer++2012}. These authors constructed phenomenological
analytical models under the epicyclic approximation with the purpose of
understanding the spiral arms. In their models gas parcels follow weakly oval
orbits, whose major axes orientation changes with radius giving rise to
kinematic density waves a-la Lindblad, and hence to spiral arms. 

An alternative viewpoint which does not consider the spiral arms as density waves has also been 
discussed in a series of papers by \cite{Romero2006,Romero2007,Athan2009b,Athan2009a, Athanassoula2010}. 
Their theory, which is more directly applicable to stars than to gas, is based on the observation that orbits in the vicinity of unstable Lagrangian points can be trapped into 
invariant manifolds whose morphology can reproduce the spiral arms.

Many authors have noted the presence of bar-driven spiral arms in
simulations, and discussed them in a more or less descriptive way \citep[see
for
example][]{Athan92b,englmaiergerhard1999,bissantzetal2003,combesrodriguez2008}.
For a recent review see also Section 2.3 of \cite{DobbsBaba2014}. Note that
in all works cited above the gas is driven by an external potential generated
by the bar and is not self-gravitating, and therefore the spiral arms are not
spiral density waves in the sense of \cite{LinShu1964}. We do not
discuss self-gravity in this work.

The viewpoint that spiral arms can be understood as density waves is 
nowadays often assumed, but has never been tested with sufficient detail.  
On modern computers, it is very cheap to run
simulations able to test the predictions and the limits of the
phenomenological models cited above.
Moreover, the literature only addresses
the weak bar regime under the epicyclic approximation, and does not discuss
how the picture should be extended to the strong bar case. 

In this paper we investigate the physical mechanism responsible for the generation of the spiral arms 
and in particular we resume the discussion of how they can be understood as kinematic density waves. 
The structure is as follows. In Sec. \ref{sec:review} we briefly review previous work aimed at understanding in a
phenomenological fashion the spiral arms in the epicyclic approximation. Then
we run grid based, isothermal, non-self gravitating 2D hydrodynamic
simulations in an externally imposed rigidly rotating barred potential,
addressing both the weak and the strong bar case. The numerical methods
employed are explained in Section \ref{sec:methods}. In Sec.
\ref{sec:testing} we compare the results of the simulations with the
phenomenological models available in the literature. In Sec.
\ref{sec:weakbar} we discuss in more detail the weak bar case, and in Sec.
\ref{sec:strongbar} how the results for the weak bar case should be extended
to the strong bar case. Finally, we summarise our findings in Sect.
\ref{sec:conclusion}.

\section{Review of previous work} \label{sec:review}
In this section we briefly review previous work aimed at understanding the
bar driven spiral arms in the epicycle approximation. In Sec. \ref{sec:epi1}
we discuss the equations describing ballistic closed orbits in the epicycle
approximation \citep[see for example Sect. 3.3.3 in][]{BT2008}. In Sec.
\ref{sec:epi2} we discuss how various authors have modified the ballistic
equations in order to describe gaseous parcels, and the phenomenological
models based on these equations. These phenomenological models will be in
later sections compared with the results of hydro simulations.

\subsection{Ballistic closed orbits in the epicycle approximation} \label{sec:epi1}
Consider a rigidly rotating external potential of the form
	\begin{equation} \label{eq:pot1}
		\Phi(R,\theta) = \Phi_0(R) + \Phi_1(R,\theta) \;,
	\end{equation}
where $\Phi_0$ is an axisymmetric potential and $\Phi_1$ is a small but otherwise arbitrary perturbation. The potential is assumed to be rigidly rotating at pattern speed $\Omega_{\rm p}$. 
$R,\theta$ are polar coordinates in the rotating frame, with $\theta=0$ corresponding to the positive horizontal axis, and $\theta$ increasing clockwise. The equation of motion for a ballistic particle in this potential is 
	\begin{equation}
		\ddot{\bfx} = - \nabla \Phi + \Omega_{\rm p} ^2 \bfx 
- 2 \Omega_{\rm p} \left( \hat{\bfe}_z \times \dot{\bfx} \right)\;, \label{eq:ballistic1}
	\end{equation} 
where $\hat{\bfe}_z$ is the unit vector perpendicular to the plane. The first term on the RHS represents gravitational forces, the second term is the centrifugal force and the third is the Coriolis force.

We want to find closed orbits in the above potential under the epicycle approximation. In the absence of the perturbation $\Phi_1$ the potential is axisymmetric and the only stable closed orbits are circular orbits. These circular orbits will be described in polar form as
	\begin{align}
		R(t) &= R_0\;, \\
		 \theta(t) &= \theta_0(t) = \Omega_{\rm f} t\;, \qquad 
	\end{align}
where 
\begin{equation}  \Omega_{\rm f} = \Omega - \Omega_{\rm p}\;, \end{equation} 
and $\Omega = \Omega(R_0)$ is the angular velocity for circular motion at
radius $R_0$ in the potential $\Phi_0$. Now consider the situation in the
presence of the small perturbation $\Phi_1$. We want orbits that are closed
in the rotating frame.\footnote{Recall that the notion of closureness of an
orbit is frame dependent. Orbits that are closed in the inertial frame are in
general not closed in the rotating frame and vice-versa.} We expect that far
from resonances closed orbits will be weak oval deformations of the circular
orbits found in $\Phi_0$. We look for closed orbits of the form
	\begin{align}
		R(t) &= R_0 + R_1(t)\;, \label{eq:p1}
		\\ \theta(t) & = \theta_0(t)  + \theta_1(t) \;,
\label{eq:p2}
	\end{align}
 where all quantities with subscript $1$ are to be considered small.  By
expressing Eq. \eqref{eq:ballistic1} in polar components, then substituting
Eqs. \eqref{eq:p1}, \eqref{eq:p2} and \eqref{eq:pot1} into it and
approximating to first order in quantities with subscript $1$, the equations
of motion take the following form:
	\begin{align}
		& \ddot{R}_1 
+ \left( \frac{\di^2 \Phi_0}{\di R^2} - \Omega^2 \right)_{R_0} R_1 - 2 R_0
\Omega \dot{\theta}_1 = - \left(\frac{\pa \Phi_1}{\pa R}
\right)_{R_0,\theta_0(t)} \;, 
 \label{eq:ballistic2}
		\\ & \ddot{\theta}_1 + 2 \Omega \frac{\dot{R}_1}{R_0}  =
		-\frac{1}{R_0^2} \left(  \frac{\pa \Phi_1}{\pa \theta}
		\right)_{R_0,\theta_0(t)} \;,
\label{eq:ballistic3}
	\end{align}
 where the derivatives of $\Phi_1$ are evaluated along the unperturbed
trajectory $[R_0,\theta_0(t)]$ and are therefore to be considered given
functions of time. Exploiting the fact that $\theta_0(t)$ is a linear
function of time, Eq. \eqref{eq:ballistic3} can be integrated immediately to
obtain $\dot{\theta}_1$, which can then substituted into Eq.
\eqref{eq:ballistic2}. This gives
	\begin{equation}
		\boxed { \ddot{R}_1 + \kappa_0^2 R_1 =  f_1(t) }, \label{eq:ballistic4}
	\end{equation}
where 
	\begin{align}
		\kappa_0^2 & =  \left( R \frac{\di\Omega^2}{\di R \,\,} + 4 \Omega^2  \right)_{R_0} \;, \\
		f_1(t) & =- \left[ \left(\frac{\pa \Phi_1}{\pa R}\right) +  \frac{2 \Omega }{\Omega_{\rm f} R_0} \Phi_1  \right]_{R_0,\theta_0(t)} \;.
	\end{align}
 Eq. \ref{eq:ballistic4} is the equation of a forced harmonic oscillator. The
natural frequency of this oscillator is $\kappa_0$, the usual epicycle
frequency that can be calculated from $\Phi_0$. The driving force is
$f_1(t)$, and as one would expect it is due to the perturbing potential
$\Phi_1$ and reduces to zero when this is turned off. $f_1(t)$ is a periodic
function of time. If we expand it in its Fourier components with respect to
time, we will find the same components contained in the Fourier expansion of
$\Phi_1$ with respect to $\theta$. This is not surprising, as the perturbed
potential encountered by the guiding centre of the ballistic particle is
given by $\Phi_1(R_0,\theta_0(t))$ where $\theta_0(t)$ increases linearly
with time.

The general solution of Eq. \eqref{eq:ballistic4} could in principle be easily written down. For each solution, we could eliminate $t$ from Eqs. \eqref{eq:p1}, \eqref{eq:p2} to find the orbit in the form $R=R(\theta)$. Not all solutions of Eq. \eqref{eq:ballistic4} correspond in general to closed orbits in the rotating frame. Some particular solutions describe closed orbits, while the others in general describe non closed loop orbits. The first type of solution are the ballistic closed orbits in the epicycle approximation. 

\begin{figure}
\includegraphics[width=0.5\textwidth]{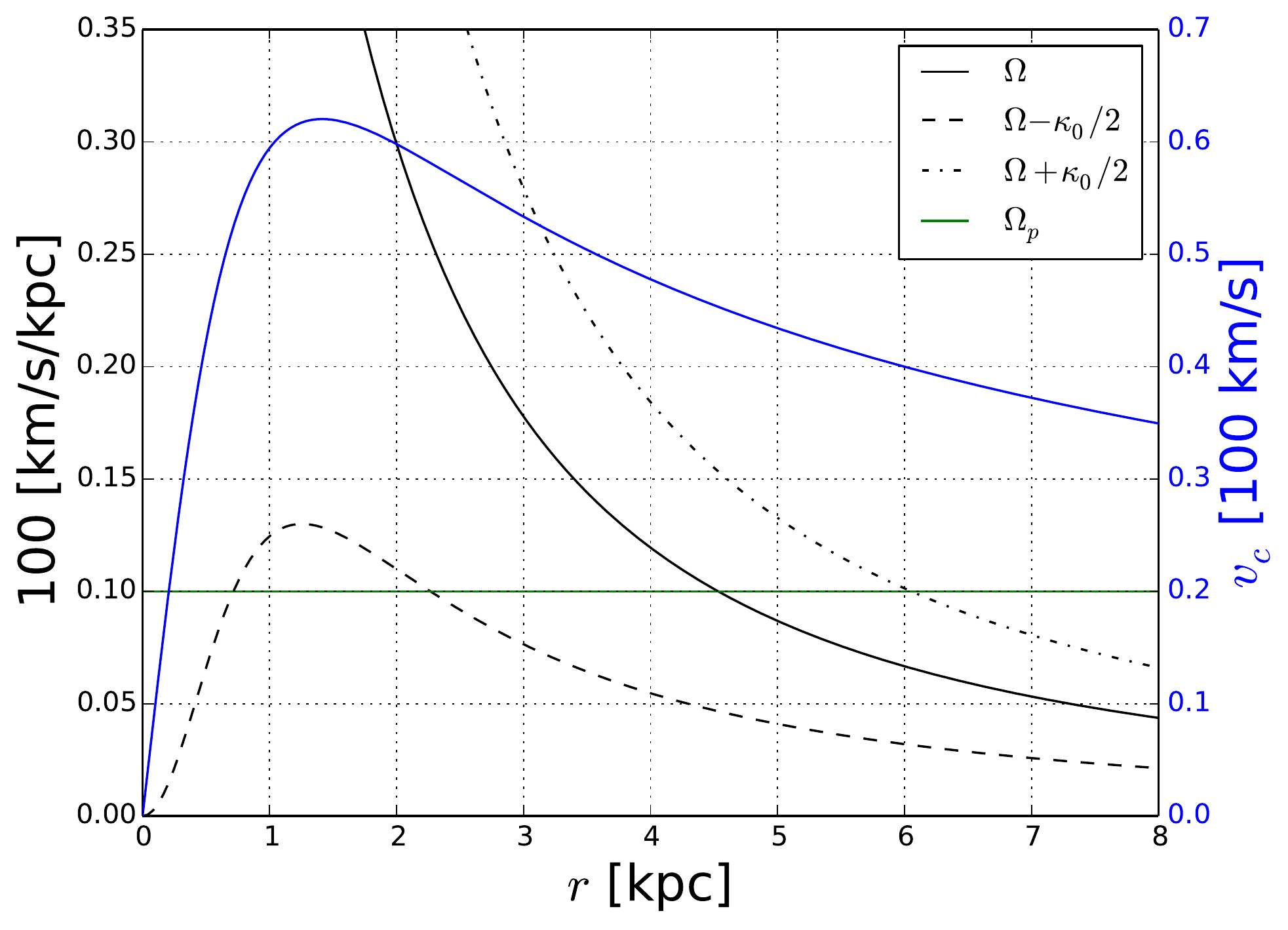}
 \caption{Blue curve: circular speed curve for the potential in Eq.
\eqref{eq:potWada} with values of parameters indicated in the text. Black
curves: behaviour of $\Omega - n \kappa_0/2$ for the same potential. The
horizontal green line indicates the assumed value of the pattern speed.}
\label{fig:wada1}
\end{figure}

For concreteness, let us now consider the particular form of the potential
used by \cite{Sanders1977} and \cite{Wada1994}:
	\begin{align} 
		& \Phi_0(R) = - \frac{ a v_0^2 }{\sqrt{R^2 + a^2}}\;, \label{eq:potWada} \\
		& \Phi_1(R,\theta) = \Phi_{\rm b}(R) \cos(2 \theta)\;, \qquad
		\Phi_{\rm b}(R) = - \epsilon \frac{  (a v_0 R)^2}{(R^2 + a^2)^2}\;, \label{eq:potWada2}
	\end{align}
 where $a$, $v_0$, $\epsilon$ are constant parameters. The axisymmetric part
$\Phi_0$ is a Kuzmin-Toomre potential, and the perturbation $\Phi_1$ was
introduced by \cite{Sanders1977}.  Following \cite{Wada1994}, we choose
$a=1\kpc$ and $v_0 = 100 \kms$ and a pattern speed of $\Omega_{\rm p}=10 \kms
\kpc^{-1}$. Different values of $\epsilon$ will be considered in the
following. Fig. \ref{fig:wada1} shows the circular speed curve and the
behaviour of $\Omega - n \kappa_0/2$, from which we can determine the
location of the resonances, for this choice of values of the parameters. 
For this potential, Eq. \eqref{eq:ballistic4} becomes
	\begin{equation}
		\boxed { \ddot{R}_1 + \kappa_0^2 R_1 =  f_{\rm w}\cos(2 \Omega_{\rm f} t) }\;, \label{eq:ballistic5}
	\end{equation}
where 
	\begin{equation}
		f_{\rm w}(R_0) = - \left[ \left(\frac{\di \Phi_{\rm b}}{\di R}\right) +  \frac{2 \Omega }{\Omega_{\rm f} R_0} \Phi_{\rm b}  \right]_{R_0} .
	\end{equation}
The forcing is now a periodic function of time with frequency $2 \Omega_{\rm f}$. By solving \eqref{eq:ballistic5}, it is found that ballistic closed orbits have the major axis always either perpendicular or parallel to the major axis of the bar.\footnote{These correspond to orbits obtained by taking the limit $\lambda \to 0$ of the solution that neglect transients of Eq. \eqref{eq:dissipation1} below, which are also the orbits with $C_1 = 0$ in the notation of Sect. 3.3.3 in \cite{BT2008}. It should be noted that other closed orbits solutions are possible at special radii, i.e. when the ratio $\kappa_0 / (\Omega - \Omega_{\rm p})$ happens to be a rational number, and correspond to orbits with $C_1\neq 0$. However these are isolated cases that are not part of a family of closed orbits parenting non closed orbits, and therefore are not of interest in this paper.} The orientation changes abruptly at each ILR, at CR, and at OLR. Note however that the above analysis is not valid in the vicinity of these points: at all Lindblad resonances the natural frequency of the oscillator $2\Omega_{\rm f}$ tends to the forcing frequency $\kappa_0$, while at CR the forcing $f_{\rm w}$ becomes infinite. The orientation of the ballistic closed orbits will be important in the next subsection.

\begin{figure}
\includegraphics[width=0.5\textwidth]{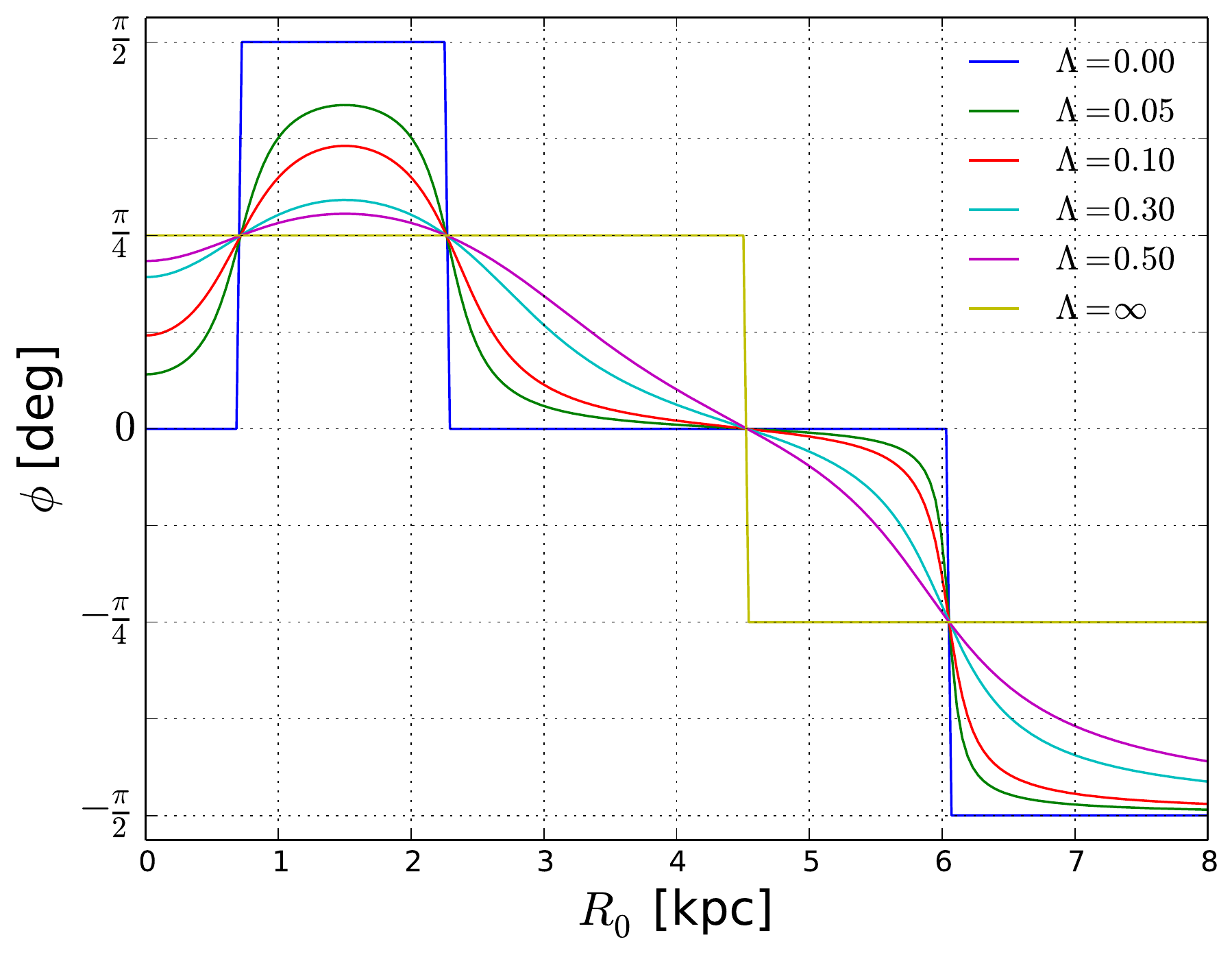}
 \caption{Predictions of Wada's model for the phase $\phi$ as function of
guiding-centre radius $R_0$ for the potential in Eq.~\eqref{eq:potWada}.
$\phi$ represents the orientation of the major axis of a closed orbit. Curves
corresponding to different values of the damping parameter $\Lambda$ are
shown. The possible values allowed by the phenomenological model are enclosed
between the blue and yellow lines.} \label{fig:wada2}
\end{figure}

\subsection{Gaseous closed orbits in the epicycle approximation: phenomenological models} \label{sec:epi2}
In the previous section we have seen that for the particular potential given
by Eqs.~\eqref{eq:potWada} and \eqref{eq:potWada2} the significant ballistic closed orbits
are always oriented either perpendicularly or parallel to the major axis of
the bar, with abrupt changes of orientation at each resonance. However, the
discussion was only valid for ballistic particles. How should we modify it to
describe the motion of a gaseous parcel which is part of a continuous fluid?
\cite{SandersHuntley1976} suggested that in the gaseous case the orientation
shift between horizontally and vertically aligned orbits happens gradually, and that
this gives rise to kinematic spiral arms a-la Lindblad. Moreover, they noted
that the addition of a dissipation term to Eq. \eqref{eq:ballistic5},
	 \begin{equation}
	 	\boxed{ \ddot{R}_1 + 2 \lambda \dot{R}_1 + \kappa_0^2 R_1 = f_{\rm w} \cos\left(2 \Omega_{\rm f} t \right)}\;, \label{eq:dissipation1}
	 \end{equation}
can remove the divergence of the radial oscillation
amplitude at the Lindblad resonaces, and that the solution of Eq.
\eqref{eq:dissipation1} where transients\footnote{By transients we mean the part of the solution that vanishes in the limit $t\to\infty$. In this case we have a driven and damped harmonic oscillator. The general solution is the sum of a decaying exponential and an oscillatory term, and the decaying exponential is the transient.} are neglected always describes a
closed orbit. They also noted that at each ILR, the closed orbit solution so
obtained has a major axis inclined at $\pi/4$ independently of the value of
$\lambda$. Since $\pi/4$ is halfway between horizontally and vertical
elongated orbits, this corroborated the conjecture that in the gaseous case
we expect a gradual shift from vertically elongated to horizontally elongated
when we cross a Lindblad resonance, which is the key to the generation of spiral arms
in their model.

\begin{figure}
\includegraphics[width=0.5\textwidth]{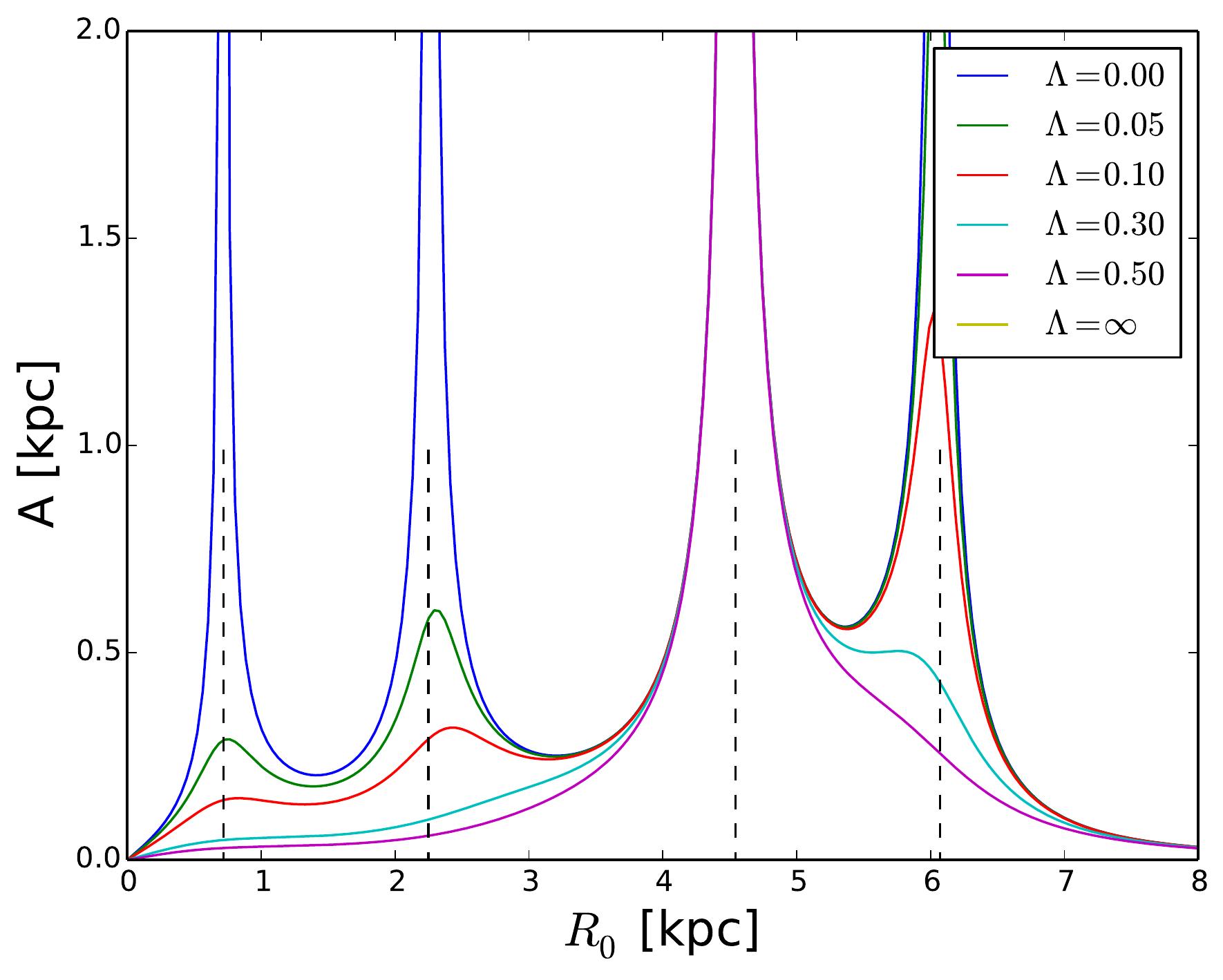}

\includegraphics[width=0.5\textwidth]{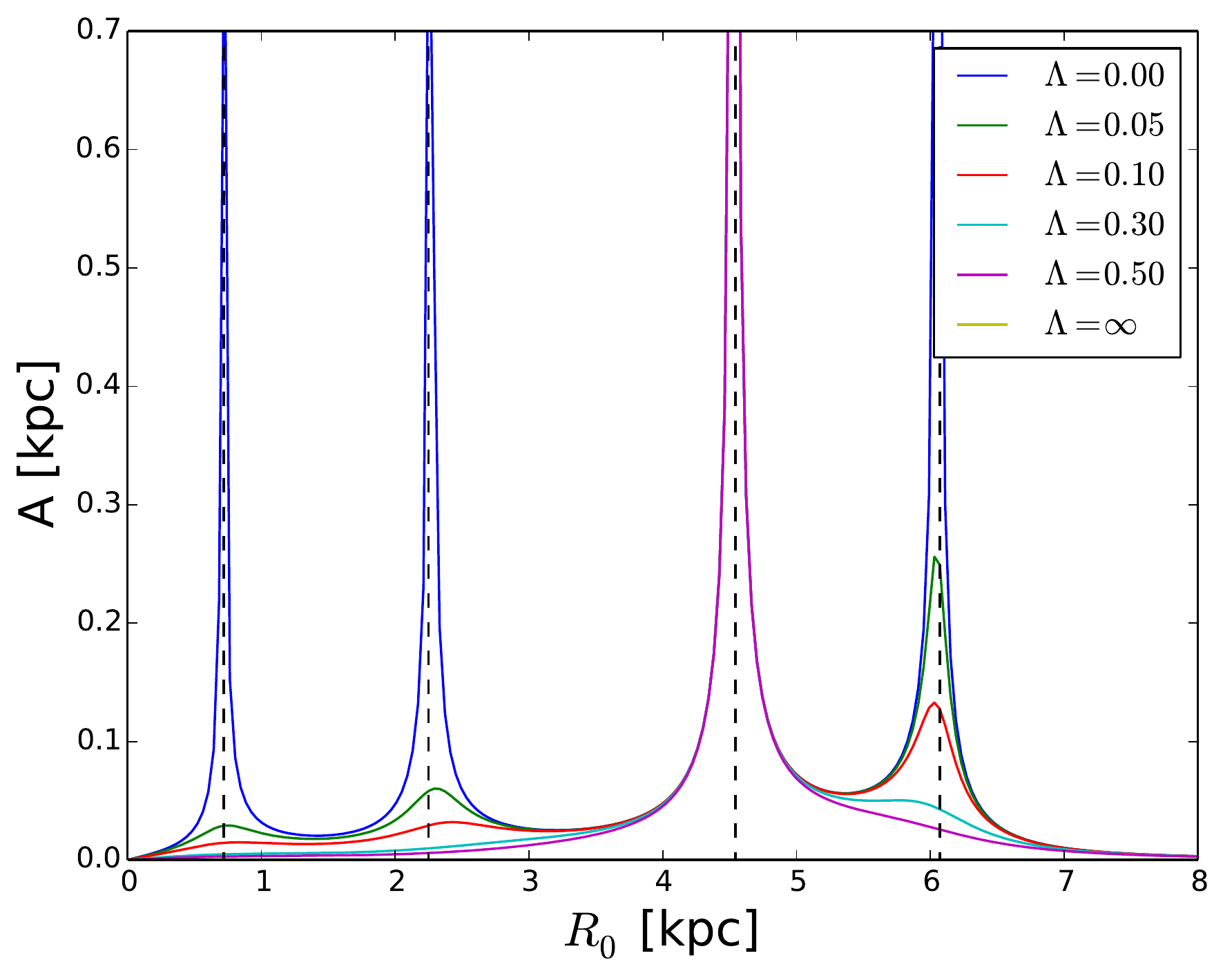}
 \caption{Amplitude of the radial oscillations predicted by {Wada's} model
for two different values of the bar strength $\epsilon$. The top panel is for
$\epsilon=0.05$, and the bottom panel is for $\epsilon=0.005$. In each panel, curves
corresponding to different values of the damping parameter $\Lambda$ are
shown. In absence of a damping term, $\Lambda=0$, the amplitude diverges at
all Lindblad resonances and at corotation. The presence of a damping term
makes the amplitude finite at the Lindlblad resonances, but not at
corotation.} \label{fig:wada3}
\end{figure}

\cite{Wada1994} and \cite{LindbladLindblad1994,PinolFerrer++2012}
independently developed further this idea by taking more seriously the
dissipation term, considering it as a phenomenological model that describes
the trajectory of gaseous particles over extended regions. The approaches
of \cite{Wada1994} and \cite{LindbladLindblad1994} differ slightly in the way
they implement the dissipation term. The phenomenological model of \cite{Wada1994} 
is based exactly on Eq. \eqref{eq:dissipation1}. This amounts to add a
dissipation term for the motion of the particle in the radial direction only.
\cite{LindbladLindblad1994} added a dissipation term that is proportional to
the difference between the velocity of the particle and the local circular
speed in the underlying axisymmetric potential. Here, we consider in more
detail the variant of \cite{Wada1994}, but the results of the
\cite{LindbladLindblad1994} variant are qualitatively similar, and a more
detailed account of the results using this variant can be found in
\cite{PinolFerrer++2012}. 

Let us now briefly review Wada's model. The solution of
Eq.~\eqref{eq:dissipation1} that excludes transient terms yields closed orbits of
the following form:
	\begin{equation}
		R(\theta) = R_0 + A \cos 2 \left(\theta - \phi\right)\;, \label{eq:epi1}
	\end{equation}
where the amplitude is
	\begin{equation}
		A = \frac{| f_{\rm w} |}{F}\;,
	\end{equation}
with
	\begin{equation}
		F = \sqrt{\left[ \kappa_0^2 - \left( 2\Omega_{\rm f}\right)^2 \right]^2  +\left(4 \lambda \Omega_{\rm f}\right)^2  }\;.
	\end{equation} 
The phase $\phi$ in~\eqref{eq:epi1} is given by the solution of
	\begin{equation}
            \begin{split}
		\sin 2 \phi &= \frac{4\lambda |  \Omega_{\rm f} |}{ F}
		\text{sign}(f_{\rm w})\; , \\\
		\cos 2 \phi &= \frac{\kappa_0^2 - \left( 2\Omega_{\rm f}\right)^2 }{F}\;, \
              \end{split}
              \label{eq:phi}
	\end{equation}
that lies in the range $-\pi \leq 2 \phi \leq \pi$.  It 
encodes the information about the orientation of the major
axis of the closed orbit, and is the inclination of the major axis of the
orbit with respect to the horizontal axis. From Eq.  \eqref{eq:phi} we see
that the quadrant $2 \phi$ belongs to is determined by the sign of $f_{\rm
w}$, which flips at CR, and by the sign of $\kappa_0^2 - 2\Omega_{\rm f}^2$,
which flips at each Lindblad resonance. The phase $\phi$ in Eq.
\eqref{eq:epi1} is determined for all orbits once $\lambda=\lambda(R_0)$ is
given at each radius. Since $\lambda$ has the dimension of a frequency, it is
convenient to express it as a multiple of a characteristic frequency at that
radius. Following \cite{Wada1994}, we express $\lambda$ as a multiple of the
epicycle frequency at that radius, namely
 \begin{equation}
\lambda= \Lambda \kappa_0\;.
 \end{equation}
Fig.~\ref{fig:wada2} reproduces Fig. 2 of \cite{Wada1994} and shows his model
predictions for the phase $\phi$ for different constant values of $\Lambda$.
In the limit $\Lambda \to 0$ we recover the ballistic case, while in the
limit $\Lambda \to \infty$ all orbits are circular. The curves $\Lambda=0$
and $\Lambda=\infty$ bound the possible values that the $\phi$ can assume in
its model. Note that this plot does not depend on the strength of the bar
potential $\epsilon$, because the phase $\phi$ does not depend on the
magnitude of $f_{\rm w}$ but only on its sign. 

$A$ is the amplitude of radial oscillations. Fig.~\ref{fig:wada3} shows the
amplitude $A$ of the radial oscillations predicted by the model for various
values of $\Lambda$ and two different values of $\epsilon$. As can be seen
from this figure, $A$ always diverges at CR, regardless of the value of
$\Lambda$, but diverges at the Lindblad resonances only in the absence of the
damping term. Note also that, for given $\epsilon$, away from
resonances the value of $A$ is limited from above: the maximum value that $A$ can reach at given $R_0$ is limited by its value for $\Lambda=0$. Finally, note that for a
given potential, this theory has only one adjustable parameter at each
radius, the damping $\lambda$. Both $\phi$ and $A$ at each radius depend on a
single phenomenological parameter in this theory. 

\begin{figure*}
\includegraphics[width=1.0\textwidth]{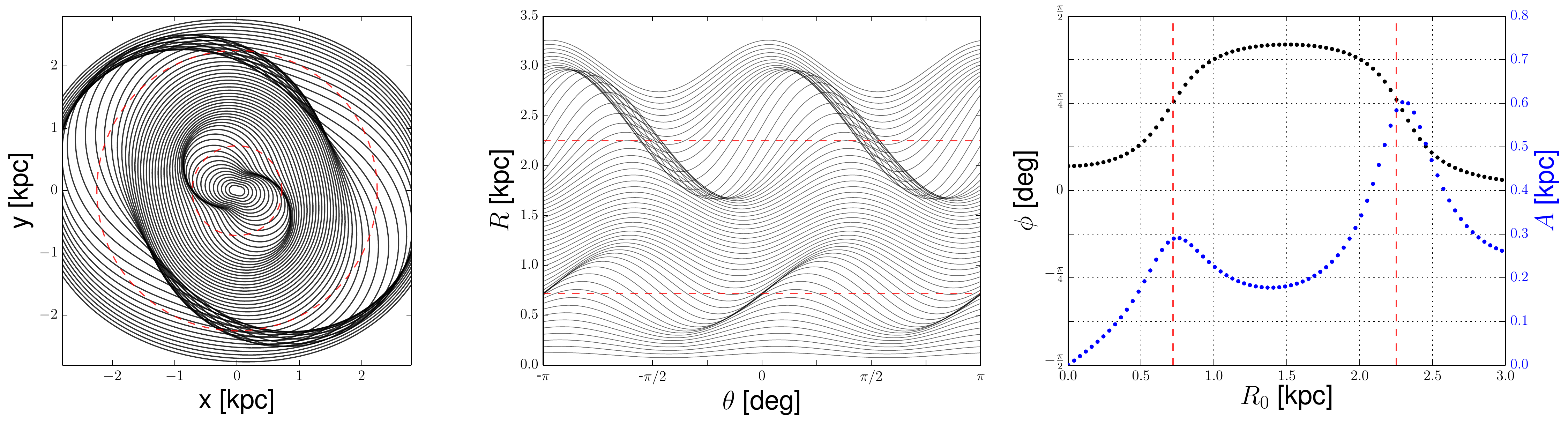}
\includegraphics[width=1.0\textwidth]{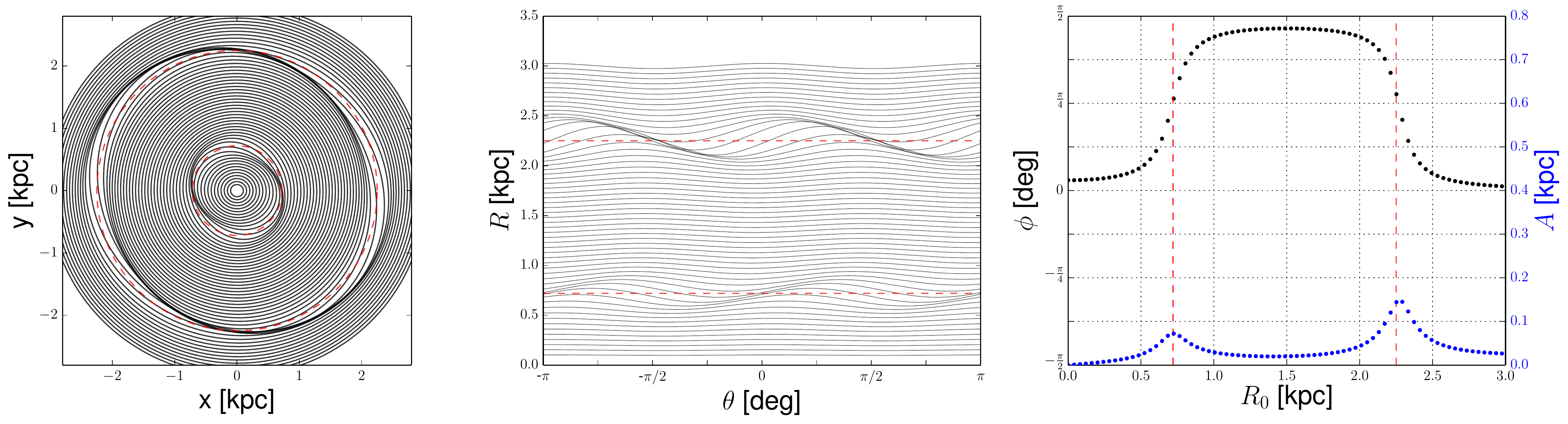}
 \caption{Top: reproduction of \protect\cite{Wada1994} predictions for a bar strength of
$\epsilon=0.05$. The damping parameter is $\Lambda=0.05$. Bottom: another
model produced using \protect\cite{Wada1994} theory, for $\epsilon=0.005$ and
$\Lambda=0.02$.} \label{fig:wada4}
\end{figure*}

Using the theory outlined above, it is possible to construct
explicit models of the spiral arms. Fig.~\ref{fig:wada4}, top panels,
reproduces the model shown in Fig. 4 of \cite{Wada1994}. It shows a nested
sequence of closed orbits for values of the parameters $\epsilon=0.05$ and
$\Lambda=0.05$. The bottom panels show another model, for a weaker bar
$\epsilon=0.005$ and $\Lambda=0.02$.

\section{Numerical methods} \label{sec:methods}

In our simulations, we assume that the gas is a fluid governed by the Euler
equations complemented by the equation of state of an isothermal ideal gas.
Then we run two-dimensional hydrodynamical simulations in an externally
imposed, rigidly rotating barred potential.  The output of each simulation
consists in snapshots of the velocity and surface density distributions
$\rho(\bfx)$ and $\bfv(\bfx)$ at chosen times.

We use a grid-based, Eulerian code based on the second-order flux-splitting
scheme developed by \cite{vanAlbada+82} and later used by \cite{Athan92b},
\cite{weinersellwood1999} and others to study gas dynamics in bar potentials.
We used the same implementation of the code as was used by \cite{SBM}, which
also implements the recycling law of \cite{Athan92b}.
 
We used a grid $N \times N$ to simulate a square $8 \, \kpc$ on a side.  $N$
depends on the resolution of the simulation.  The initial conditions of the
runs are as follows.  We start with gas in equilibrium on circular orbits in
an axisymmetrized potential and, to avoid transients, turn on the
non-axisymmetric part of the potential gradually.  We use outflow boundary
conditions: gas can freely escape the simulated region, after which it is
lost forever. The potential well is sufficiently deep, however, to prevent
excessive quantities of material to escape. 

The recycling law introduces a term in the continuity equation that was
originally meant to take into account in a simple way the effects of star
formation and stellar mass loss.  The equation governing this process is
\begin{equation}\label{eq:recycle}
\frac{\p \rho}{\p t} =\alpha (\rho_0^2 - \rho^2)\;,
\end{equation}
where $\alpha=0.3 \, M_\odot \pc^{-2}\Gyr^{-1}$ is a constant and $\rho_0$ is
the initial surface density, which is taken to be $\rho_0 = 1 \, M_\odot
\pc^{-2}$.  In practice, the only effect of the recycling law is to prevent
too much gas accumulating in the very centre and to replace gas lost at the
boundary due to the outflow boundary conditions. It does not affect the
morphology of the results, so our results do not change if we disable the
recycling law. 
\section{Phenomenological Models vs Hydro simulations} \label{sec:testing}

\begin{figure*}
\includegraphics[width=1.0\textwidth]{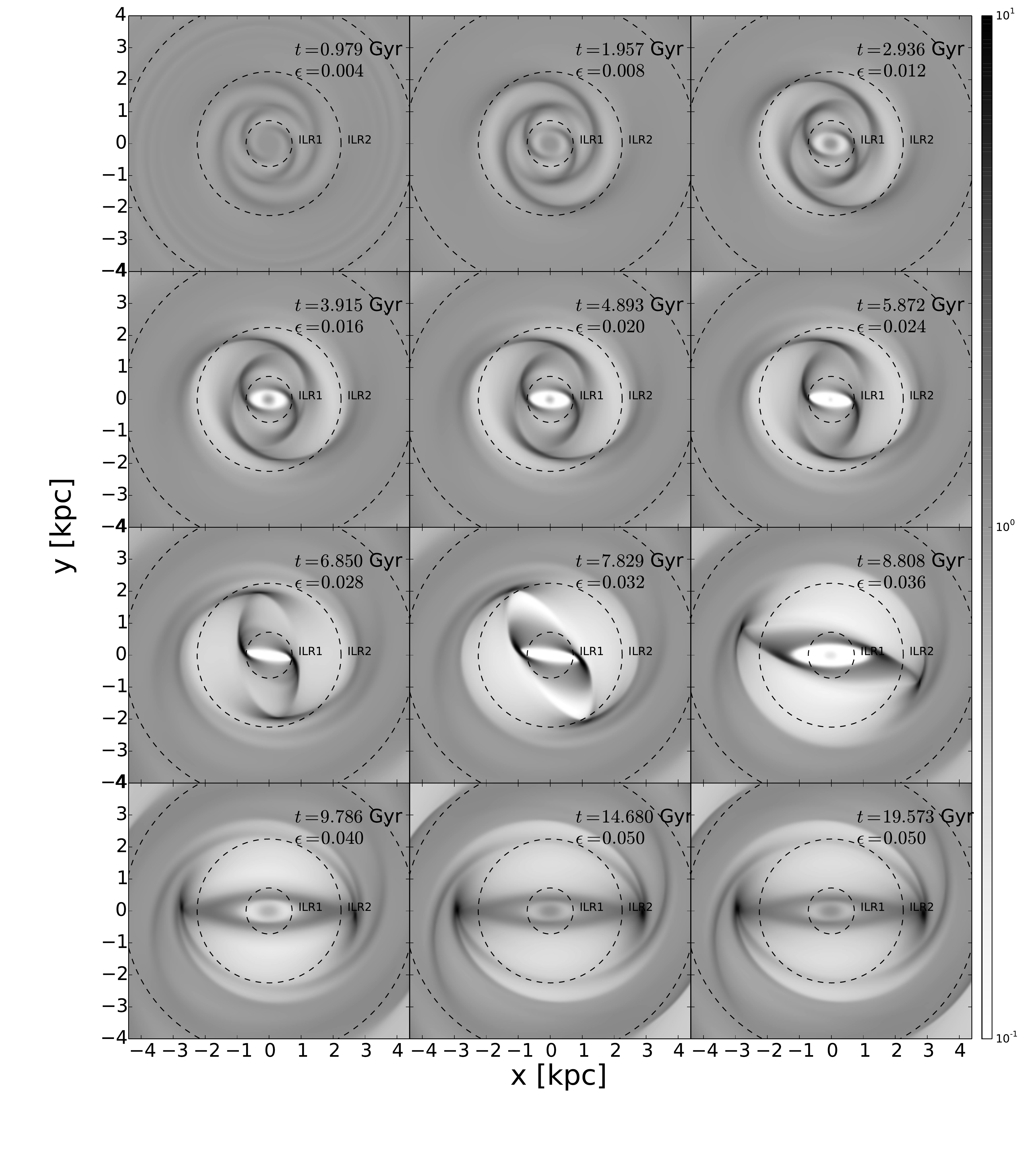}
 \caption{Time sequence for a simulation in the potential defined by
Eqs.~\eqref{eq:potWada} and \eqref{eq:potWada2}. Each panel shows the density map at a particular time. The bar strength $\epsilon$
increases slowly and linearly with time until it reaches the value of
$\epsilon=0.05$, and is then frozen. The gas adjusts almost
adiabatically to the instantaneous underlying potential. The sound speed is
$c_s = 2.5 \kms$ and the spatial resolution is $\di x = 20 \pc$. The dotted
circles indicate the two ILRs and the CR. In each panel, the evolutionary time $t$
and the instantaneous value of $\epsilon$ are shown. The colorbar is in units of $M_\odot \pc^{-2}$}
\label{fig:wadasequence} 
\end{figure*}

What happens if we run a simulations in the potential of Eqs.
\eqref{eq:potWada} and \eqref{eq:potWada2}? Do the results of a hydro
simulation resemble the predictions of the phenomenological models?
Fig.~\ref{fig:wadasequence} answers this question. It shows the results of a
simulation in which $\epsilon(t)$ is a slowly changing function of time. At
$t=0$ we have $\epsilon=0$, then $\epsilon$ increases linearly with time
until it reaches Wada's value of $\epsilon=0.05$ at $t=12.3\rm Gyr$, and for
later times it is kept constant at this value. The potential evolves so
slowly that the gas flow configuration evolves almost adiabatically, adjusted
to the instantaneous underlying potential. In other words, at each time $t$
the gas configuration is almost the same as the steady state configuration
that would be obtained by freezing the potential and then waiting for the gas
to settle down into a steady state. 

At early times, when $\epsilon$ is very small, the results of the hydro
simulation do qualitatively resemble the predictions of the phenomenological
models, as can be seen by comparing the top row in
Fig.~\ref{fig:wadasequence} with Fig.~\ref{fig:wada4}. At later times, when
$\epsilon$ is greater, this is not true: for example panels in the bottom row
do not look like \cite{Wada1994} predictions. Inspection of streamlines shows
that for very small values of $\epsilon$ the gas is flowing on almost
circular orbits, as the epicycle approximation requires, but when $\epsilon$
becomes larger the gas flows instead on very horizontally elongated orbits,
and the epicycle approximation is not valid anymore. How can we explain this
behaviour? Why does the gas flow on very elongated orbits despite the fact
that the perturbation is apparently very small? As we shall now see, the key
to answer these questions lies in the orbital structure of the underlying
potential. 

\begin{figure*}
\includegraphics[width=1.0\textwidth]{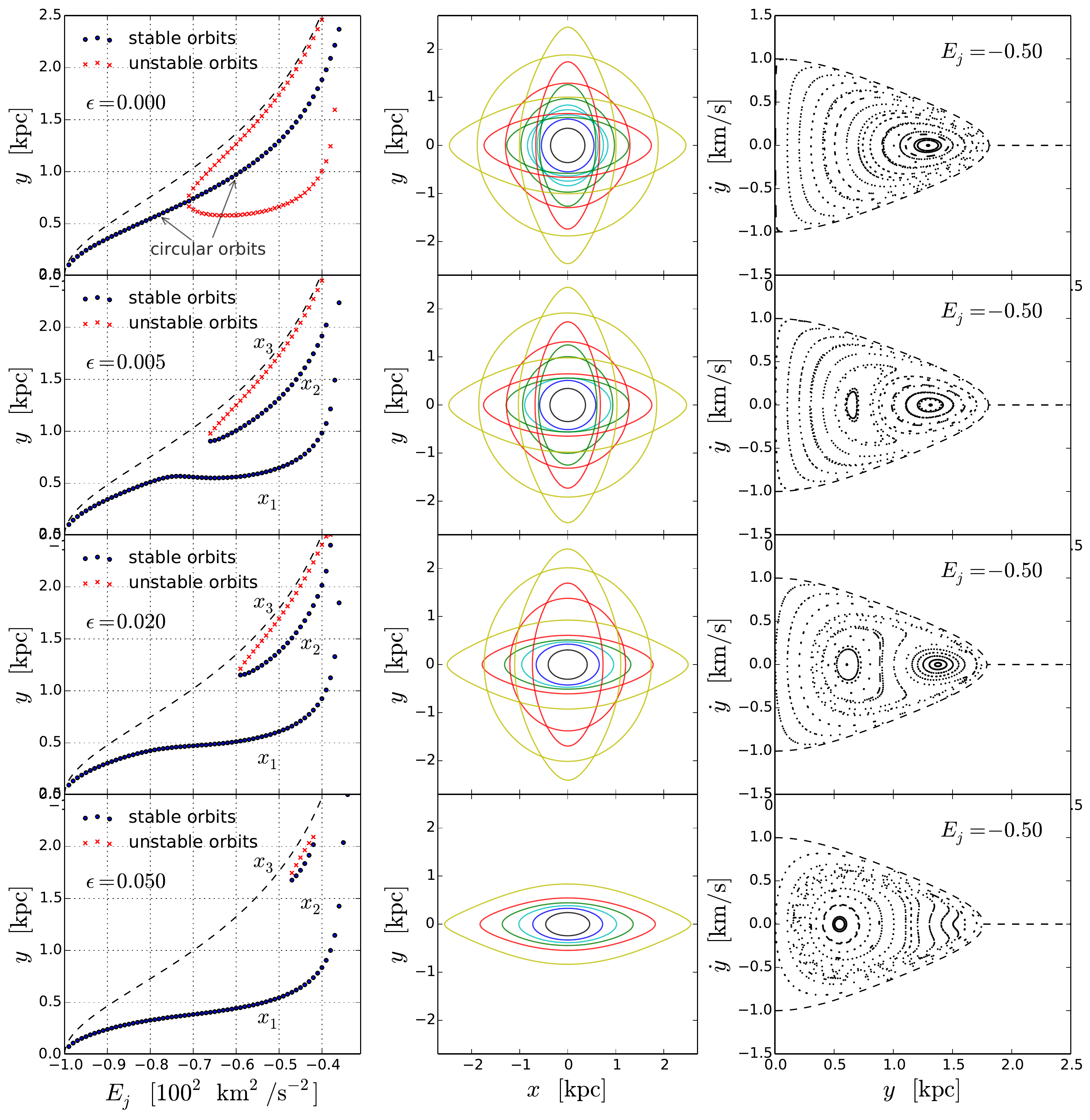}
 \caption{Orbital structure of the potential given by Eq.~\eqref{eq:potWada}
for different values of $\epsilon$. Rows from top to bottom show the cases
$\epsilon= 0.0, 0.005, 0.02, 0.05$, respectively. In the left column, the
Jacobi constant $E_j$ of a closed orbit is plotted against the value at which
the orbit cuts the vertical axis in potentials with different values of
$\epsilon$. Each dot represents a closed orbit, and all closed orbits cut the
vertical axis with purely horizontal velocity. The middle column shows the
closed orbits in $xy$ space, with orbits of the same colour having the same
energy. The orbits shown are those corresponding to values of $E_j$ equal to
$\{-0.9,-0.8,-0.7,-0.6,-0.5,-0.4\}$. The vertical slices of the grid in the left panels are drawn at these energies. 
The correspondence between energy and colours can be identified from the bottom panel in the central column: here greater energy (i.e., less negative) 
corresponds to bigger orbits, and the same colour coding applies to other panels. The right column shows surfaces of section
for  $E_j=-0.5$.}
\label{fig:wadaorbits1}
\end{figure*}

Fig.~\ref{fig:wadaorbits1} shows how the orbital structure of the underlying
potential changes as $\epsilon$ is varied. Each row refers to a particular
value of $\epsilon$. Let us first consider the top row, which shows the case
$\epsilon=0$, when the bar perturbation is turned off and the potential is
exactly axisymmetric. Consequently, orbits that close in our rotating frame
are either circular or they are orbits for which the precession frequency
happens to coincide with our chosen ``pattern speed'', which is actually of
no physical significance when $\epsilon=0$. The top left panel of
Fig.~\ref{fig:wadaorbits1} shows several families of closed orbits: each dot
represents a closed orbit in terms of its value of the Jacobi constant $E_j$
and the coordinate at which the orbit cuts the vertical axis. All the orbits
shown in this diagram \citep[known as characteristic diagram, see][]{contopoulosreview} have the property of cutting the vertical axis at right
angles.  The line of dark dots shows the circular orbits.  Since for
$\epsilon=0$ the potential is axisymmetric, these are the only stable closed
orbits. 

The red crosses in the top left panel of Fig.~\ref{fig:wadaorbits1} show
eccentric orbits that close in the rotating frame. Such orbits exist only
between the two ILRs. They spring out from the sequence of nearly circular
orbits at a radius of the innermost inner Lindblad resonance ILR1, and then
merge into it again at a bigger radius corresponding to the other inner
Lindblad resonance, ILR2. The red crosses that are above the dark line are 
elongated perpendicularly to the bar, while those below it are elongated along the bar. 
In fact, since the potential is axisymmetric, the red cross above the dark line and
that vertically below it represent orbits that differ only in a $90\degree$
rotation of the major axis.  Since the potential is axisymmetric, it is
actually possible to find equivalent orbits for any orientation, but these
are not shown in the diagram as they do not cut the vertical axis at right
angles. The central panel of Fig.~\ref{fig:wadaorbits1} shows
how the orbits appear in the $xy$ plane. These orbits are coloured differently according to their value of the Jacobi constant $E_{j}$.

We now consider what happens when a small perturbation is turned on, so the
potential is no longer exactly axisymmetric. The left panel of the second row
of Fig.~\ref{fig:wadaorbits1} shows that for $\epsilon=0.005$, a bifurcation
is present approximately at the location of ILR1. The line of dark points
that marks the circular orbits in the axisymmetric case is now split. The
part inside ILR1 merged with what used to be the orbits elongated along the bar in the axisymmetric case (which have become stable) to form a
continuous line.  In fact, this continuous line constitutes the $x_1$ family
in the notation of \cite{contopoulosreview}.
The part that used to be circular orbits between the two ILRs now forms a closed loop with the
orbits elongated perpendicularly to the bar. In fact, the former circular orbits between the
ILRs have become the stable $x_2$ family, while the orbits elongated perpendicularly to the bar 
have become the $x_3$ family and have remained unstable, as opposed to
the $x_1$ orbits that have become stable.  Note also that the $x_3$ orbits
are not the $x_1$ orbits rotated as in the axisymmetric case, because the
potential is no longer axisymmetric.

When in the bottom two rows of Fig.~\ref{fig:wadaorbits1} we further increase
$\epsilon$, the $x_2$-$x_3$ loop shrinks, and has almost disappeared when
$\epsilon=0.05$. Note that the shape of the orbits, visible in the central
column, does change as we increase $\epsilon$, but not dramatically.  
What changes significantly is the fraction of the volume of
phase space that is occupied by non-closed orbits that librate around $x_1$
orbits rather than around $x_2$ orbits. The right column of
Fig.~\ref{fig:wadaorbits1} illustrates this point by showing surfaces of
section\footnote{However, one must be careful not to confuse the full
phase-space volume occupied by a group of orbits with the area in a surface
of section filled by the same orbits, see \cite{Binney++1985}.} for a value
of Jacobi constant $E_j=-0.5$ \citep[for a definition of surfaces of section see for example Chapter 3 in][]{BT2008}. When $\epsilon=0$, all non closed orbits are
parented by the circular orbit, which corresponds to the centre of the
``eye'' in the top-right panel. When $\epsilon=0.005$, two eyes are present
in the surfaces of section; the centre of the left one corresponds to the
$x_1$ orbit, the centre of the right one to the $x_2$ orbit (which replaces
the circular orbit). Some orbits are now parented by the $x_1$ orbit, and
some others by the $x_2$. As we increase $\epsilon$, the $x_1$ orbits becomes
predominant in the surface of section, and the fraction of orbits parented by
it increases until the $x_2$ orbit disappears for this energy. 
 
It is now easy to go back to Fig.~\ref{fig:wadasequence} and interpret the
results. When $\epsilon$ is small, the gas is circulating on $x_2$ orbits in
the outer parts and on weakly elongated $x_1$ orbits in the inner part
(inside the ILR1).\footnote{Note that there must be a transition region in between, 
and spiral arms are generated as the orientation of orbits changes to make this transition. When the bar perturbation is stronger and orbits are more elongated, the transition can be mediated by shocks \citep[see][]{SBM} rather than by a soft spiral arm overdensity as in this case. Note also that in the present case the transition is $x_2 \to x_1$ outwards-in, while the transition discussed in \cite{SBM} is $x_1\to x_2$.}
When $\epsilon$ is increased, the volume of phase space
occupied by orbits parented by $x_2$ orbits is reduced. $x_2$ orbits start
disappearing at small radii, and as they disappear the gas has no other
choice than to settle onto $x_1$ orbits. When
$\epsilon=0.05$ the $x_2$ orbits are gone almost everywhere, and the gas is
flowing everywhere on $x_1$ orbits, including the region where the $x_1$
orbits are highly elongated. In this regime, the epicycle approximation
obviously cannot work as the $x_1$ orbits are not weakly deformed circular
orbits. If we call weak a bar that can be well described under the epicycle
approximation, then the $\epsilon=0.05$ case should be classified as a strong
bar, despite the fact that the non-axisymmetric part of the potential might
naively appear small compared to the axisymmetric part.

We have now two questions. 
\begin{enumerate}
\item How do the phenomenological theories reviewed above compare with the
hydro simulations in the weak bar case, that is, when $\epsilon$ is extremely
small and the epicycle approximation is valid?

\item What can we say in the strong bar case, when the gas is flowing on very
elongated orbits and the epicycle approximation is not valid?

\end{enumerate}
 Sections \ref{sec:weakbar} and \ref{sec:strongbar} investigate these two
questions respectively.

\section{Weak bar Case} \label{sec:weakbar}

In this section we want to investigate how the phenomenological models
reviewed above compare in detail with the results of a hydro simulation in
the weak bar case, when the epicycle approximation is valid. To do this, we
study the case $\epsilon=0.005$. Note that this is a value ten times smaller
than the value $\epsilon=0.05$ considered by \cite{Wada1994} and
\cite{PinolFerrer++2012}. As we have seen in the last section, that should
actually be considered a strong bar case, because of its orbital structure. 

\begin{figure*}
\includegraphics[width=1.0\textwidth]{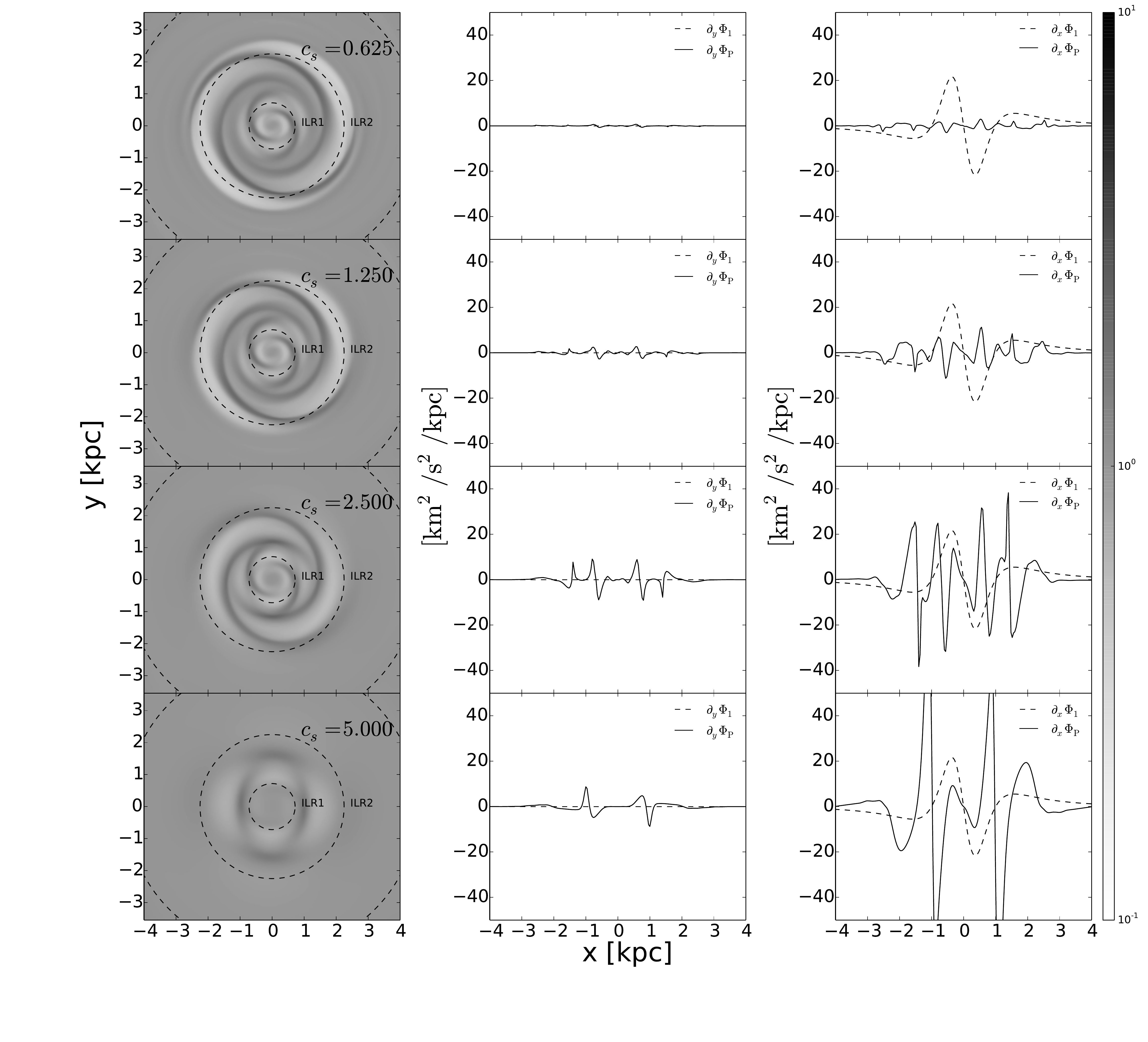}
 \caption{Results of a simulation in {Wada's} potential in the case of a weak
bar ($\epsilon=0.005$) for different values of the sound speed $c_s$. Each
panel in the left column shows the steady-state density long after the bar has reached the final value of $\epsilon=0.005$ -- compare with
Fig.~\ref{fig:wada4}. The middle column shows the $y$ components of pressure
forces (full lines) and bar perturbation gravitational forces (dashed line) from
the potential in Eq. \ref{eq:potWada} with $\epsilon=0.005$
calculated along the horizontal axis. The right column shows the $x$
components of the same forces, again calculated along the horizontal axis.
Note that pressure forces can be derived from a pressure potential $\Phi_P$
as described in the text.} \label{fig:weakbar1}
\end{figure*}

When the sound speed is not too high, the morphology of the density
distribution qualitatively resembles the predictions of \cite{Wada1994} shown
in Fig. \ref{fig:wada4}. This can be seen from Fig. \ref{fig:weakbar1}, which
shows results of hydro simulations with $\epsilon=0.005$ and different values
of the sound speed $c_s$. In these simulations the bar strength is frozen after the value $\epsilon=0.005$ is reached,
and the snapshots considered are all at $t=2.9 \rm Gyr$, long after the bar is fully grown to this value 
and the gas has long settled down into a steady-state configuration. The left column shows the density distribution. The density
distributions for $c_s=0.625\kms$ and $c_s=1.25\kms$ are very similar: it
appears as if the gas configuration is tending towards some limiting
configuration as $c_s\to0$ at fixed spatial resolution, and that this
limiting configuration has much in common with the predictions of the
phenomenological models.

\begin{figure*}
\includegraphics[width=1.0\textwidth]{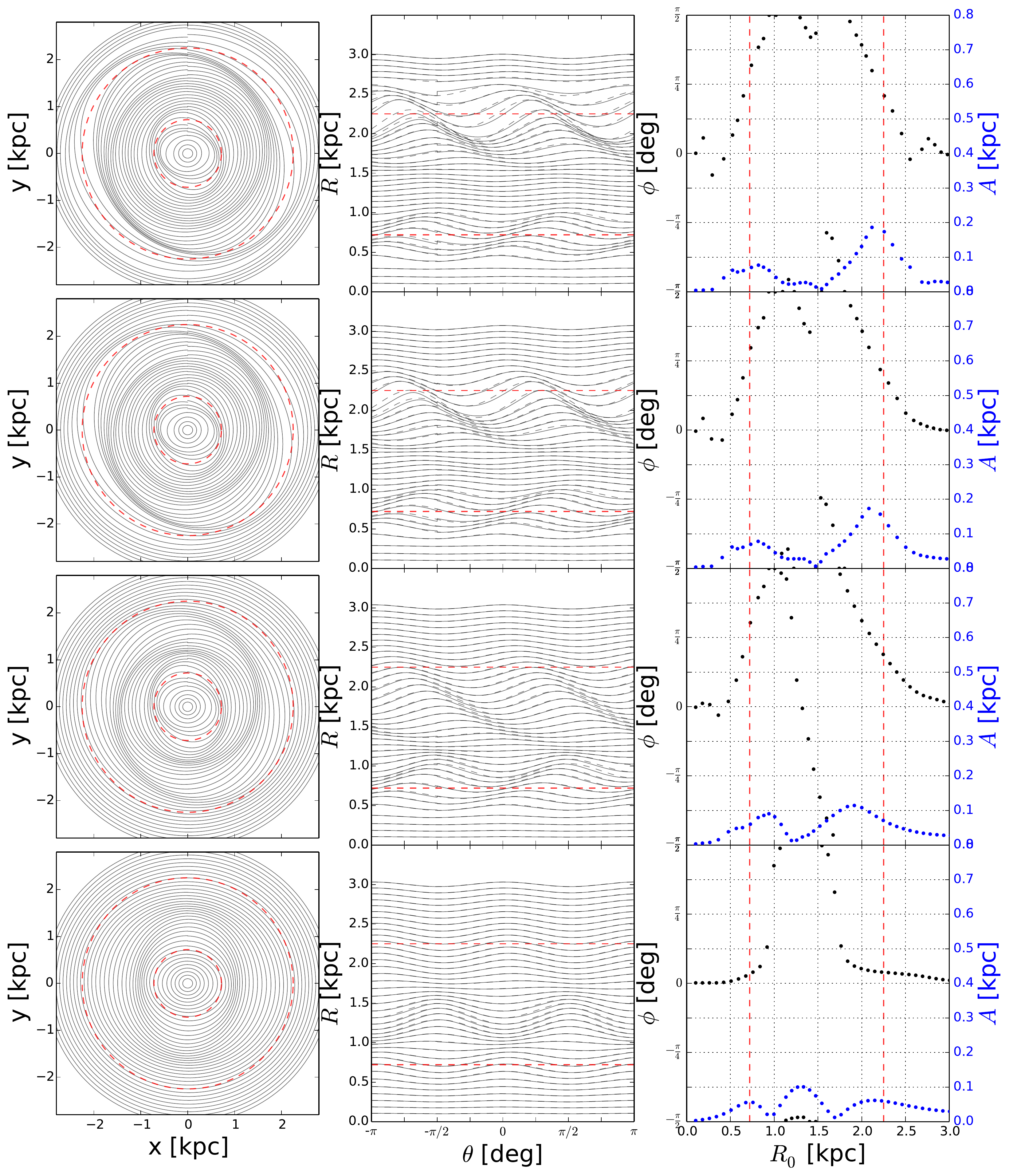}
 \caption{Analysis of the instantaneous streamlines for the snapshots shown
in Fig. \ref{fig:weakbar1}. The left column shows a sequence of nested
streamlines in the $xy$ plane. The middle column shows in full lines the same
streamlines in the $R\theta$ plane, and in dashed lines the result of a
simple fit using Eq.~\ref{eq:ballistic4}, where $A$ and $\phi$ are
independent parameters. The right panel shows the best fit values of $A$ and
$\phi$ as a function of radius. In each panel the dashed red lines mark the
two inner Lindblad resonances.} \label{fig:weakbar2}
\end{figure*}

A second important prediction of the phenomenological model that is confirmed
by the hydro simulations is that the spiral arms can be understood as
kinematic density waves. Fig. \ref{fig:weakbar2} investigates the
instantaneous streamlines, that is, the streamlines calculated from a frozen
snapshot of the velocity field. Since the gas is in a steady state, these are
not much different from the real streamlines followed by gas parcels during
the time evolution. We see that in all cases streamlines are almost
closed curves, in the sense that in general the mismatch after a single loop
is much less than the extent of a radial oscillation during the loop. 

Another verified prediction is that streamlines are well described by closed
orbits of the form Eq. \eqref{eq:epi1}. The middle column in Fig.
\ref{fig:weakbar2} shows in full lines the same streamlines as the left
column in the $R\theta$ plane, and in dashed lines a simple fit to each
streamline using Eq. \eqref{eq:epi1} where $A$ and $\phi$ are considered
independent free parameters. The streamlines are well fitted by a function of
this form, and if we were to reproduce a nested sequence of streamlines as
in Fig. \ref{fig:wada4} using the best fitting values of these two
parameters, we would obtain kinematic density waves that reproduce those
obtained in the simulation. 

However, in the phenomenological models, $A$ and $\phi$ both depend on a
single parameter, the dissipation $\lambda$, while in the fitting procedure
above $A$ and $\phi$ are treated as independent parameters. Moreover, in the
phenomenological models some values of $A$ and $\phi$ are forbidden: $\phi$
must stay within the blue and yellow line in Fig. \ref{fig:wada2}, and the
amplitude $A$ away from resonance has an upper bound. Therefore it is not
obvious that dissipation parameter $\lambda$ required by the phenomenological
model can be tuned for each streamline as to produce the values of $A$ and
$\phi$ given by a hydro simulation. In other words, the values obtained from
a hydro simulation for $A$ and $\phi$ need not to come from the same value of
$\lambda$, or could be outside the regions obtainable for any value of $\lambda$. 
Indeed, from the right column of Fig. \ref{fig:weakbar2} it is
clear that in the hydro simulation $A$ and $\phi$ are often outside the regions
allowed by the phenomenological models for the same potential. Therefore not
all hydro simulations can be reproduced by an appropriate gauging of the
dissipation. In fact, we chose to fit using $A$ and $\phi$ as two independent parameters instead of using directly 
the dissipation $\lambda$ as the only free parameter because $A$ and $\phi$ lie outside the allowed regions too often. 
However, In the limit of very low sound speed, the results of
the hydro simulations for $A$ and $\phi$ are not far from the predictions of
the phenomenological model in Fig. \ref{fig:wada4}. 
The values of $\phi$ in the top two rows of Fig. \ref{fig:weakbar2} are more similar than those in other rows 
to the black dots in the bottom right panel of Fig. \ref{fig:wada4}, and the values of $A$ 
 have also much in common with the blue dots in the same panel, although many are still forbidden because 
they lie just above the curve $A=0$ in the bottom panel of Fig. \ref{fig:wada3}.
Another interesting prediction that is verified in this limit is that the orientation of the
major axis of the closed orbits at the two ILRs is $\pi/4$. 

Why are the results of the phenomenological models well reproduced in the
limit of vanishing sound speed, and less well when the sound speed is higher?
To understand this, consider the equations of motion of a gaseous parcel
flowing in the hydro simulation. If the fluid were really an ideal isothermal
fluid, these would be
	\begin{equation} \frac{D \bfv}{Dt} = -\nabla \Phi - c_s^2
\frac{\nabla \rho}{\rho}\;, \label{eq:euler1} \end{equation}
 where $D$ indicates the material derivative, the first term on the right
hand side is the gravitational force and the second is the pressure force.
However, in a real hydrodynamical simulation at fixed spatial resolution,
some unavoidable amount of numerical viscosity, which is not included in Eq.
\eqref{eq:euler1}, will always be present. This numerical viscosity decreases
as we increase the resolution, and tends to zero as the resolution goes to
infinity. The equations of motion of a parcel of gas are therefore something
like 
\begin{equation} \frac{D \bfv}{Dt} = -\nabla \Phi - c_s^2 \frac{\nabla
\rho}{\rho} - F_{\di x}\;,
 \label{eq:euler2} \end{equation}
 where $F_{\di x}$ indicates the force arising from the finite
resolution, which we will loosely refer to as the viscous force. Hence, a
parcel in a hydro simulation is subject to three different forces:
gravitational, pressure and viscosity. 

The phenomenological models completely neglect pressure forces. Instead, they
only account for the gravitational forces and, phenomenologically, for the
viscous forces $F_{\di x}$. Note that the pressure force can be derived from
a ``pressure potential''
	\begin{equation} \Phi_P = c_s^2 \log(\rho/\rho_0)\;, \label{eq:PhiP}
\end{equation} 
where $\rho_0$ is an arbitrary number. In a steady-state
configuration, the density does not change with time, and the pressure forces
can be derived from a static potential. Hence, a gaseous parcel can be
considered to  move in a potential that is the sum of the gravitational plus
the pressure potential. This static pressure potential is not easily
derived a priori, and we can only calculate it once we are given the steady-state
configuration of the gas after solving the hydrodynamical equations.

The central and right columns in Fig. \ref{fig:weakbar1} show the pressure
and perturbation potential forces along the horizontal axis. The central
column shows forces in the $y$ direction; the gravitational forces in this
direction are zero for the chosen form of the perturbation potential. The
right column shows forces in the horizontal direction. We see that at high
sound speed, when the phenomenological models are less accurate, the
magnitude of the pressure forces is in general greater than the magnitude of
the perturbation potential forces. At $c_s=1.25$ the pressure contribution
becomes smaller than the bar perturbation contribution, and at $c_s=0.625$
the pressure forces are negligible compared to the bar perturbation
contribution. This is why the phenomenological models above work in this
limit. The phenomenological models have terms that take into account both the
effects of viscosity and  the gravitational potential,
but they do not take into account the effects of pressure. In the limit
where pressure is negligible, they work. If in the phenomenological models we
could replace the gravitational potential with the sum of the gravitational
plus the pressure one, then they would describe the results of the
simulations at finite pressure. The problem is that in general the pressure
potential is not easily obtained a priori. 

\begin{figure*}
\includegraphics[width=1.0\textwidth]{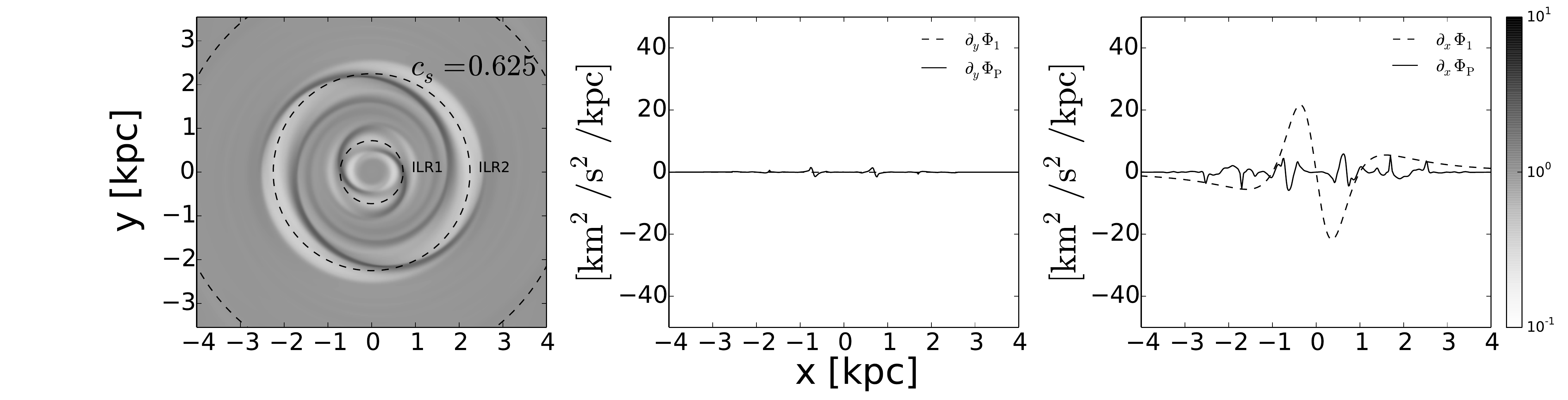}
 \caption{Same as the top row of Fig.~\ref{fig:weakbar1}, but for a
simulation with twice the spatial resolution. Here, $\di x = 10\pc$.}
\label{fig:weakbar3}
\end{figure*}

Before moving on to discuss the strong bar case, we have a couple of remarks
left. From a theoretical point of view, when the sound speed tends to zero
the gas is always supersonic and we expect all small perturbations to cause
shocks. Hence, we expect that when $c_s$ tends to zero, we would find shocks
anywhere there is a small velocity gradient. At a shock, the gradient of the
density diverges but the forces due to pressure could remain finite
as they depend on the product of the sound speed (which goes to zero) with
the gradient of the pressure (which goes to infinity). The reason why shocks
are not formed in our simulations when the sound speed tends to zero is that
at finite resolution the numerical viscosity prevents them from forming. In
the limit of vanishing sound speed, the numerical viscosity becomes more
important than the pressure force. Indeed, we have checked that increasing
the numerical resolution leads to steady states in which there are sharper
density variations -- see Fig. \ref{fig:weakbar3}. Note that the pressure
forces in Fig.~\ref{fig:weakbar3} are stronger than the pressure forces for a
simulation in the same potential, with the same value of the sound speed but
at lower resolution -- top row of Fig.~ \ref{fig:weakbar1}. Hence, we conclude
that the phenomenological models work in reproducing the result of a
simulation in the limit of low sound speed and of finite numerical viscosity,
which means finite numerical resolution.

\section{Strong Bar case} \label{sec:strongbar}

In the previous section, we have seen that in the weak bar case the spiral
arms can be understood as kinematic density waves, generated by weakly oval
closed orbits. What can we say about the strong bar case? In order to
investigate a realistic example of a strong bar we consider a simulation we
used in \cite{SBM}, performed in the \cite{binneyetal1991} barred potential.
The simulation used here is the one with sound speed $c_s=10\kms$ and spatial
resolution $\di x=10\pc$.

In \cite{SBM}, spiral arms were present and noted, but not discussed in
detail. It was shown that in the region where spiral arms are present the
velocity field was very well approximated by $x_1$ orbits. To discuss the
spiral arms, we will need to look at the tiny deviations of the hydro
velocity field from the ballistic $x_1$ orbits field. 

\begin{figure*}
\includegraphics[width=1.0\textwidth]{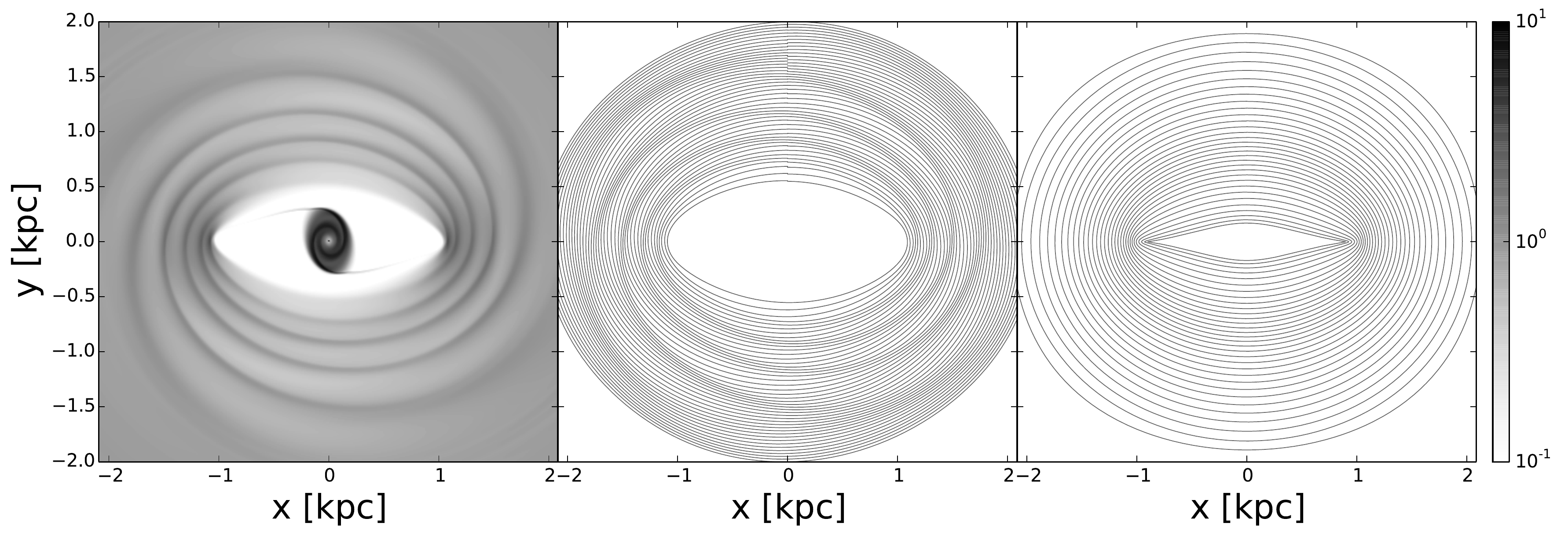}
 \caption{In the left panel, the density distribution for a simulation from
SBM2015 with $c_s=10\kms$ and spatial resolution $\di x =10\pc$. 
This snapshot is taken after the bar is fully grown and the final steady state is reached. 
The middle panel shows a nested sequence of instantaneous streamlines. The right panel
shows ballistic closed $x_1$ orbits in the same underlying potential.}
\label{fig:strongbar1}
\end{figure*}

\begin{figure}
\includegraphics[width=0.47\textwidth]{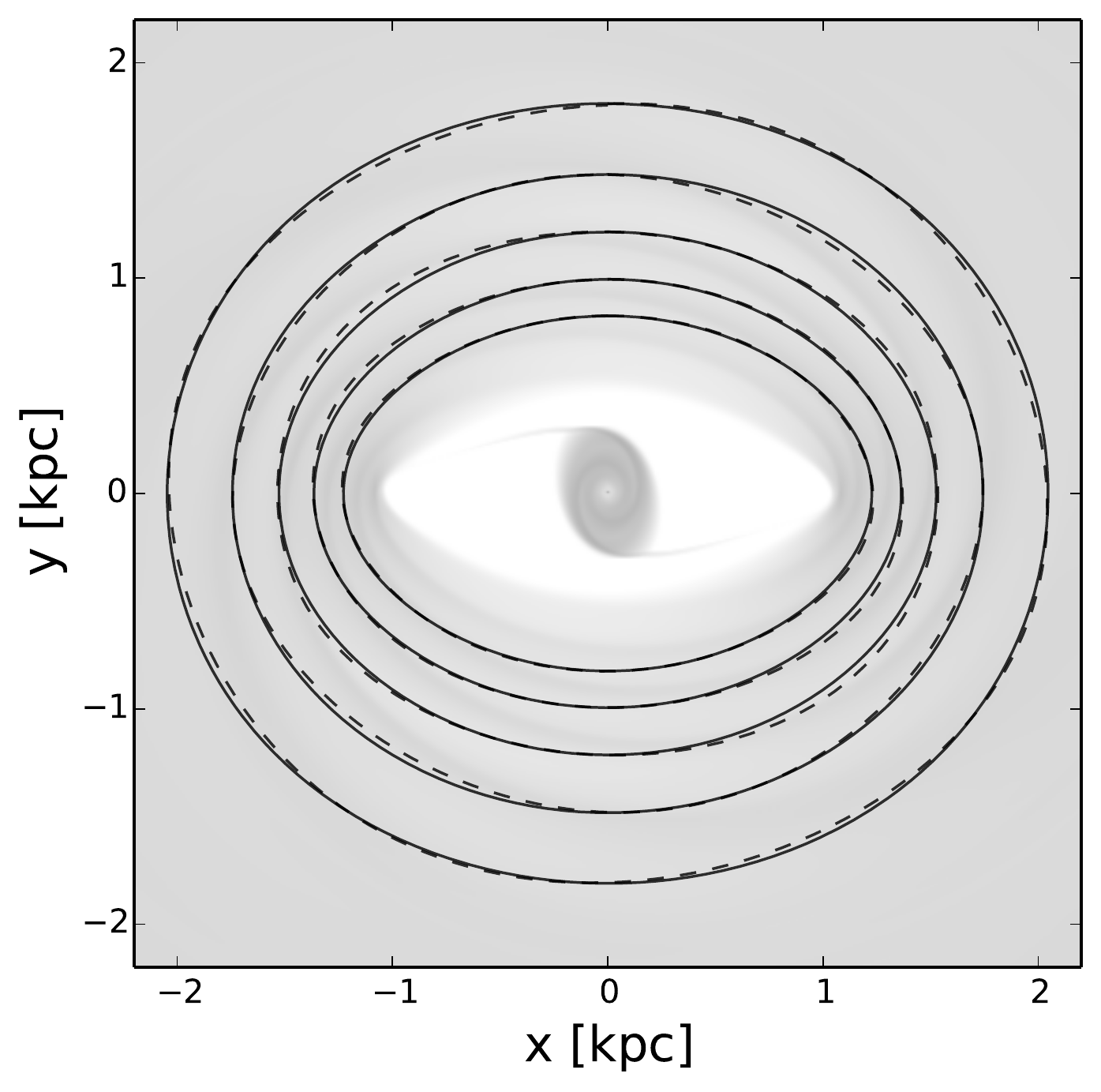}

\includegraphics[width=0.47\textwidth]{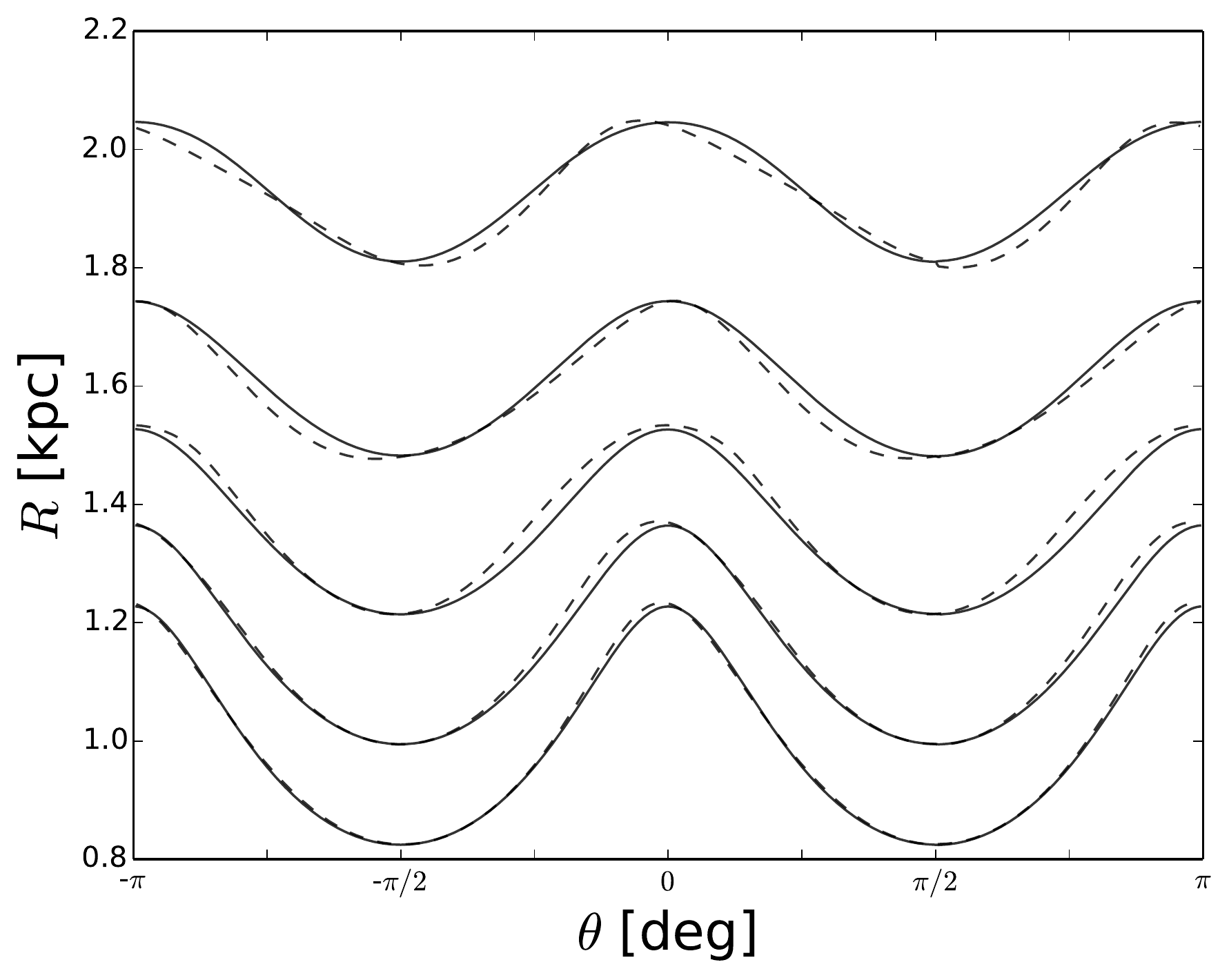}

\includegraphics[width=0.47\textwidth]{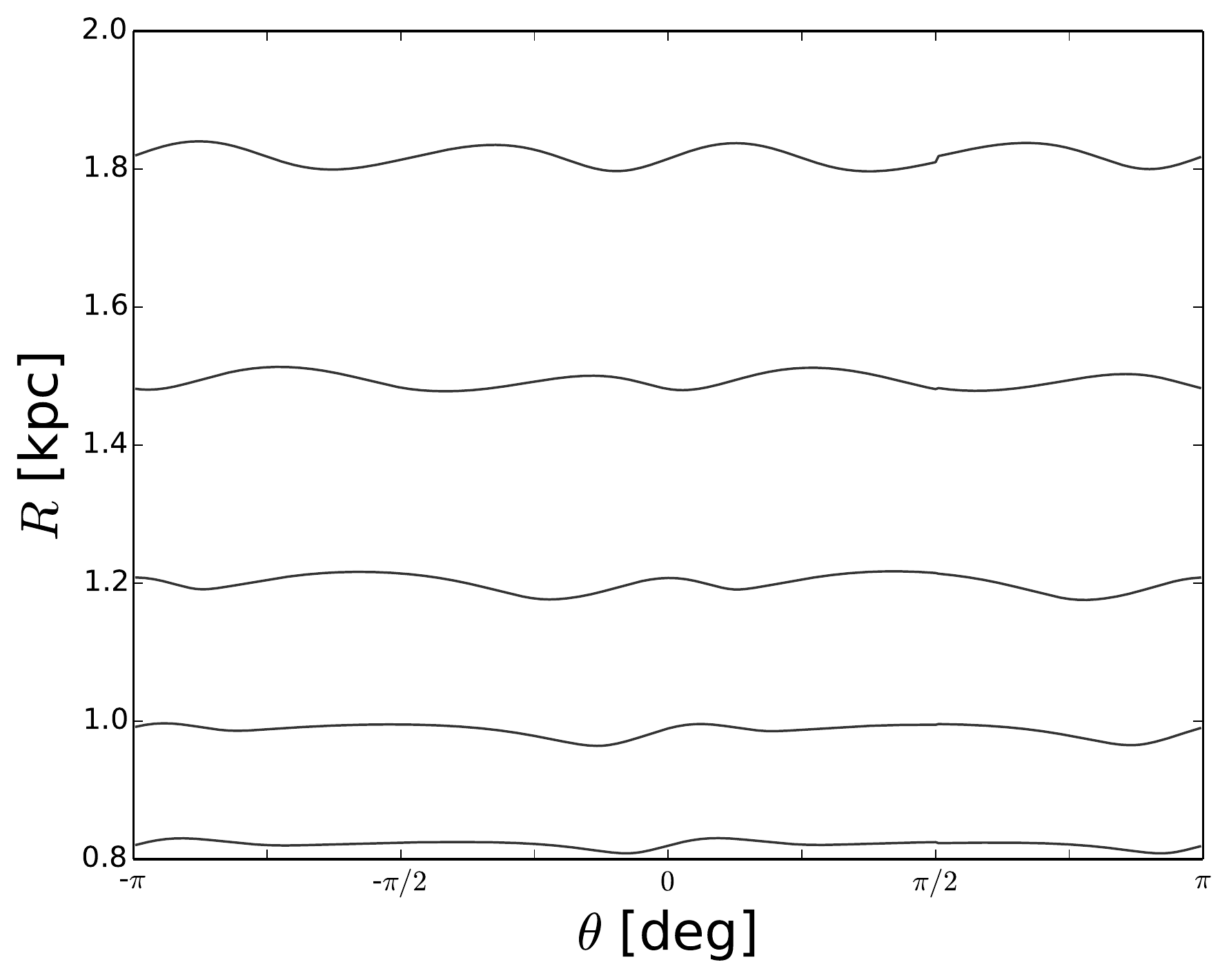}
 \caption{A Closer look at some of the streamlines of
Fig.~\ref{fig:strongbar1}. In the top panel, streamlines (dashed lines) and
ballistic $x_1$ orbits that cut the $y$ axis at the same value (full lines).
In background is visible the density distribution of
Fig.~\ref{fig:strongbar1} for ease of comparison. The middle panel shows the
same streamlines and orbits in the $R\theta$ plane. The bottom panel shows
the difference $R_{\rm streamline}(\theta) - R_{x_1}(\theta)$, where the zero
is shifted for clarity.} \label{fig:strongbar2}
\end{figure}

The left panel in Fig. \ref{fig:strongbar1} shows a density snapshot from the
simulation in \cite{SBM}, taken after the bar is fully grown and the gas has reached an approximate
steady state.  Spiral arms emerging from the end of the bar are clearly
identifiable. These spiral arms are stationary in the frame corotating with
the bar. The central panel of Fig.~\ref{fig:strongbar1} shows what
happens when we nest together many instantaneous streamlines. We see that, to
a very good approximation, streamlines are closed curves that when nested
together generate the spiral arms. In the right panel ballistic closed orbits
are shown for comparison. Fig. \ref{fig:strongbar2} analyses in more detail 
a selection of instantaneous streamlines in this snapshot. In the top panel, five streamlines are shown in
dashed lines. These are very nearly closed. In the same panel, the ballistic
closed orbits that cut the $y$ axis at the same value of the streamlines are
shown in full lines. It can be seen that streamlines are librations around
underlying closed $x_1$ orbits. In the middle panel, the same streamlines and
orbits are shown in the $R\theta$ plane. In the bottom panel, the radial
difference between the two orbits is shown in the $R\theta$ plane. The librations do
not have the simple $\cos(2\theta)$ structure that we encountered in Sect.
\ref{sec:weakbar}. 

The above considerations strongly suggest that also in the strong bar case
the spiral arms can be understood as kinematic density waves, generated by
small librations around underlying closed orbits. In the epicycle
approximation, the librations are around circular orbits. Here, the gas
parcels librate around $x_1$ orbits. It is natural to ask whether the
phenomenological models can be extended to the strong bar case, where now the
perturbations should be made around the $x_1$ orbits and not around circular
orbits. In order to investigate this, let us go back to the equations of
motion in the rotating frame. For a gaseous parcel in a hydro simulation, Eq.
\eqref{eq:ballistic1} is modified to
	\begin{equation}
		\ddot{\bfx} = - \nabla (\Phi + \Phi_P) + \Omega_{\rm p} ^2 \bfx - 2 \Omega_{\rm p} \left( \hat{\bfe}_z \times \dot{\bfx} \right)   + F_{\di x}\;, \label{eq:strongbar1}
	\end{equation} 
 where $\Phi_P$ is the ``potential'' that gives the pressure force
introduced above and $F_{\di x}$ is the viscous force. We can do a Floquet
analysis to investigate librations around ballistic closed orbits \citep[see
for example][]{BT1st}. Let us write $\Phi=\Phi_0 + \Phi_1$, where this time
$\Phi_0$ is not necessarily axisymmetric. Let $\bfx_c(t)$ be a ballistic
closed orbit in the potential $\Phi_0$. We can write a libration around this
closed orbit as $\bfx(t) = \bfx_c(t) + \bfx_1(t)$. By substituting this last
equation into \eqref{eq:strongbar1}, expanding to first order in quantities
with subscript 1 and cancelling the zeroth order terms, we obtain that the
equation of motion for the libration is 
	\begin{equation}\begin{split}
		\ddot{\bfx}_1 &= - \left[ (\bfx_1 \cdot \nabla) \nabla
		\Phi_0 +  \nabla \Phi_1 + \nabla \Phi_P  \right]_{x_c(t)} +
		\Omega_{\rm p} ^2 \bfx_1  \\
&\qquad - 2 \Omega_{\rm p} \left( \hat{\bfe}_z \times \dot{\bfx}_1 \right) + F_{\di x}\;.  \label{eq:strongbar2}
	\end{split}\end{equation} 
In this equation, all the derivatives of the potentials are to be calculated
along the unperturbed closed orbit and are therefore known functions of
time. Equations \eqref{eq:ballistic3} for the epicycle approximation can
be considered as a particular case of Eq.~\eqref{eq:strongbar2}, obtained
assuming that $\Phi_0$ is axisymmetric, $\Phi_P=0$ and there are no viscosity
forces (which are then added heuristically in the phenomenological models).
In trying to generalise the analysis of Sect. \ref{sec:review} to the case
when $\Phi$ is a strongly barred potential, we are faced with the difficulty
that there is in general no natural way of splitting $\Phi$ into a $\Phi_0$
where we can easily calculate closed orbits and a perturbation $\Phi_1$ that
only weakly deforms closed orbits. The only natural choice is to set
$\Phi_1=0$. If this choice is made, the whole perturbing potential is given
by $\Phi_P$, which is not known a priori. In the phenomenological models in
the epicycle approximation we were able to neglect $\Phi_P$ compared to
$\Phi_1$ in the low pressure case, but here the same cannot be done if
$\Phi_1=0$. 

Suppose that $\Phi_P$ is somehow known. If we assume $\Phi_1=0$ and
define a suitable phenomenological viscous term $F_{\di x}$ in Eq.
\eqref{eq:strongbar2}, would the resulting equation correctly describe
librations that give rise to spiral arms? To investigate this, we can do as
follows. We first extract the density from the snapshot shown in the left panel
of Fig. \ref{fig:strongbar1}. From this density, we derive the pressure
potential using Eq. \eqref{eq:PhiP} and $c_s = 10\kms$. 
Then we pretend we do not know how $\Phi_P$ was obtained and find librations using the following equation:
	\begin{equation} \boxed{
		\ddot{\bfx}_1 = - \left[ (\bfx_1 \cdot \nabla) \nabla \Phi_0
		+ \nabla \Phi_P  \right]_{x_c(t)} + \Omega_{\rm p} ^2 \bfx_1
		- 2 \Omega_{\rm p} \left( \hat{\bfe}_z \times \dot{\bfx}_1 \right) - 2 \lambda \dot{\bfx}_1 }\;.\label{eq:strongbar3}
	\end{equation} 
In this last equation, we have introduced a phenomenological dissipation
term analogous to the models in the epicycle approximation. The dissipation
is proportional to the difference between the total velocity and the local
$x_1$ velocity field. $\Phi_0$, $\Phi_P$ and $\Omega_{\rm p}$ are now known, and
given a value of $\lambda$ we can solve Eq. \eqref{eq:strongbar3} to find the
librations. Among all possible solutions to this equation, we want the
solutions $\bfx_1(t)$ that are periodic with the same period of the
underlying closed orbits. This amounts to finding the solution where
transients are gone, dissipated away. The method that we used to solve the
differential equation \eqref{eq:strongbar3} is described in detail in
Appendix \ref{sec:app1}. 

\begin{figure}
\includegraphics[width=0.47\textwidth]{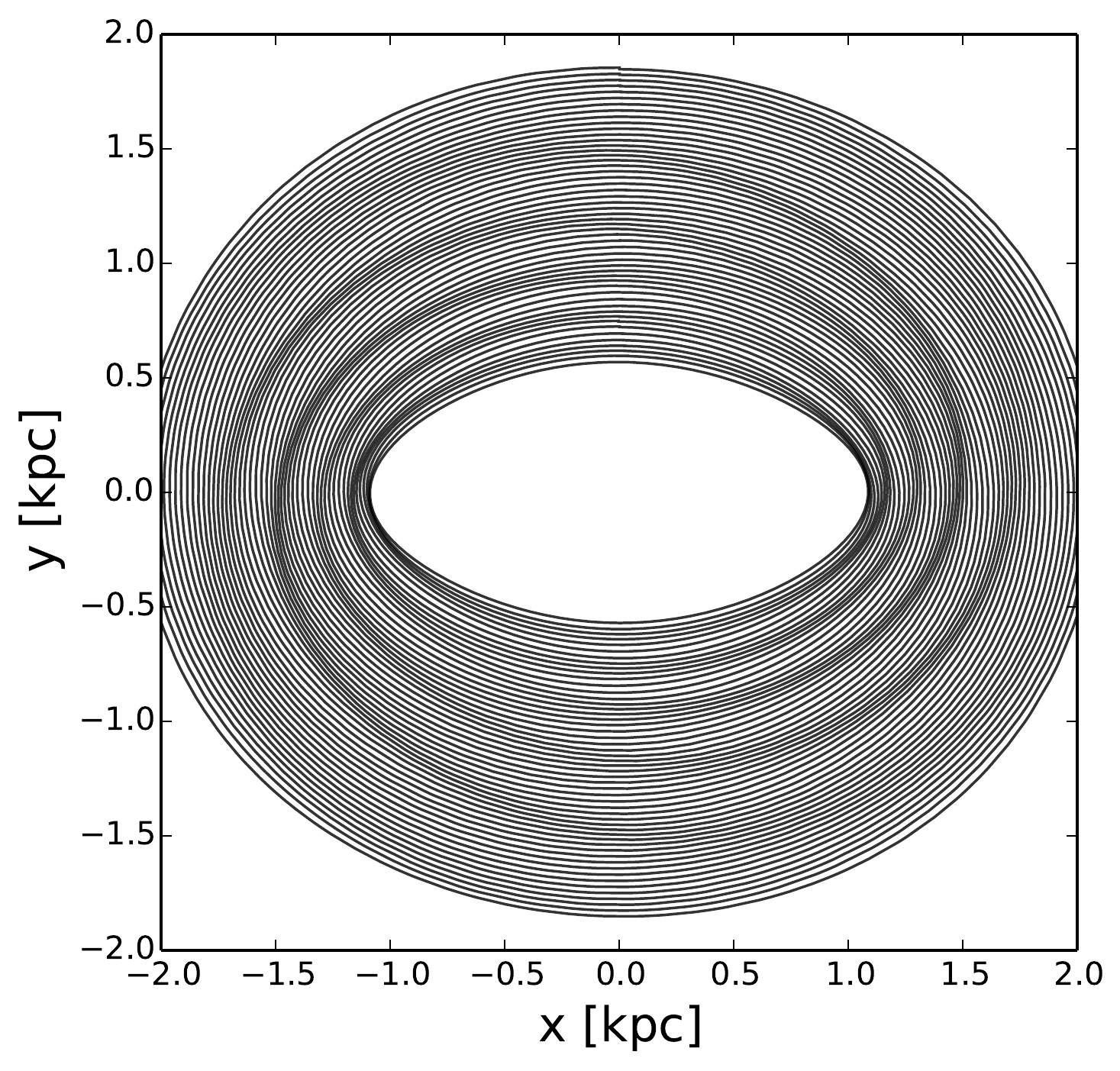}
\caption{Nested librations around closed $x_1$ orbits generated using Eq.
\eqref{eq:strongbar3}. The potential $\Phi_0$ is the potential used in
SBM2015, the pressure potential $\Phi_P$ is derived from the density map in
the left panel of \ref{fig:strongbar1}, and $\lambda=50\kms\kpc^{-1}$.}
\label{fig:strongbar3}
\end{figure}

Fig.~\ref{fig:strongbar3} shows a nested sequence of librations around $x_1$
orbits calculated using Eq. \eqref{eq:strongbar3} and a value of $\lambda =
50 \kms \kpc^{-1}$. Spiral arms identical to those present in the hydro
simulation of Fig. \eqref{fig:strongbar1} are recovered. This confirms that
the spiral arms can be understood as kinematic density waves also in the
strong bar case. Eq. \eqref{eq:strongbar3} well describes the librations if
the correct $\Phi_P$ is provided. 

However, in the above analysis we have cheated in the sense that we have
obtained $\Phi_P$ from the result a full hydrodynamical calculation. Is there
a simple way of deriving a suitable $\Phi_P$, without solving the full
hydrodynaical problem? We haven't found a better way. We considered the
possibility of obtaining $\Phi_P$ from the underlying closed orbits. Since
the $x_1$ orbits are highly elongated, we have tried to infer a density
map from orbit crowding, and from this the pressure potential. However, the
pressure forces obtained in this way were much weaker than the one obtained in
the hydro simulation for a value of the sound speed of $c_s=10\kms$, and were
not capable of reproducing spiral arms even for a small value of the
dissipation. For a higher value of the sound speed we obtained segments of
spiral arms that were not coherent and did not produce any ``grand design''.
Hence, we conclude that while the spiral arms in the strong bar case can
still be explained by kinematic density waves generated by librations around
the appropriate closed orbits, calculation of these librations require
knowledge of the pressure forces and hence solution of the full
hydrodynamical problem. Indeed, the morphology of the spiral arms found in
\cite{SBM} depends significantly on the value of pressure, and a
phenomenological model that wants to reproduce must take pressure explicitly
into account.

\section{conclusion} \label{sec:conclusion}
In this paper, we have investigated bar-driven spiral arms in the absence of
self-gravity. Our focus was on understanding the physical mechanism involved.
We concluded that, both in the weak and strong bar cases, the spiral arms can
be understood as kinematic density waves generated by librations around
underlying ballistic closed orbits. In the weak bar case, the librations can
be considered deviations from circular orbits. In the strong bar case, the
librations are to be considered deviations from the appropriate underlying
closed orbits, which can be highly elongated. In the strong bar case the
epicycle approximation is not valid. In fact, whether a bar can be considered
weak or strong is determined by the validity of the epicycle approximation.
In Sect.  \ref{sec:testing} we argued that a bar is weak or strong according
as the orbital structure is or is not obtainable from the epicycle
approximation.  Bars that might naively be considered weak, for example the
case $\epsilon=0.05$, should be considered strong. 

A parcel in a hydro simulation is subject to three different forces: gravitational, pressure and viscosity. 
We have tested the phenomenological models available in the literature aimed
at explaining the spiral arms in the weak bar case, when the epicycle
approximation is valid. We found that the key ingredient not taken into
account by these models is pressure. Therefore they work well in regimes where
pressure can be neglected, such as simulations in a weak bar at finite
resolution and vanishing sound speed. We have also discussed how the
phenomenological models should be extended to the strong bar case. When the
pressure forces are known, these extensions work very well in explaining the
spiral arms in the strong bar case. Unfortunately, the pressure forces are in
general known only after solving the full hydrodynamical problem and are not
known a priori. Thus, while the phenomenological models provide insight into
the physical mechanism that generates the spiral arms, they appear to be of
little practical use.

\section*{Acknowledgements}

MCS acknowledges the support of the Clarendon Scholarship Fund and is
indebted to Steven N. Shore for helpful discussions.  
JB and JM were supported by Science and Technology Facilities Council by grants R22138/GA001 and
ST/K00106X/1. JM acknowledges support from the ``Research in Paris''
programme of Ville de Paris.  The research leading to these results has received funding from
the European Research Council under the European Union's Seventh Framework
Programme (FP7/2007-2013) / ERC grant agreement no.\ 321067.

\def\aap{A\&A}\def\aj{AJ}\def\apj{ApJ}\def\mnras{MNRAS}\def\araa{ARA\&A}\def\aapr{Astronomy \&
  Astrophysics Review}\def\apjs{ApJS}
\bibliographystyle{mn2e}
\bibliography{2d}

\clearpage

\appendix
\section{Solving the Floquet Equation} \label{sec:app1}
In this Appendix we show how to solve Eq.~\eqref{eq:strongbar3}. This
equation arose during a Floquet analysis to find librations around closed
orbits in a strongly barred potential. The equation to be solved is:
	\begin{equation}
		\ddot{\bfx}_1 = - \left[ (\bfx_1 \cdot \nabla) \nabla \Phi_0 +  \nabla \Phi_P  \right]_{x_c(t)} + \Omega_p ^2 \bfx_1 - 2 \Omega_p \left( \hat{e}_z \times \dot{\bfx}_1 \right) - 2 \lambda \dot{\bfx}  \;.			\label{eq:sb1}
	\end{equation} 
Expanding into components, rearranging and renaming we can rewrite this equation as the following system:
	\begin{equation}\begin{split}
		\ddot{x}_1 & = - x_1 \kappa_{xx}(t) - y_1 \kappa_{xy}(t) +
		F_{x}(t)  + \Omega_p ^2 x_1 -  2 \Omega_p \dot{y}_1 - 2
		\lambda \dot{x}_1\;, \\
		\ddot{y}_1 & = - x_1 \kappa_{xy}(t) - y_1 \kappa_{yy}(t) + F_{y}(t)  + \Omega_p ^2 y_1 + 2 \Omega_p \dot{x}_1 - 2 \lambda \dot{y}_1 \;,
	\end{split}\label{eq:sb2} \end{equation}
where 
	\begin{equation}\begin{split}
	 	\kappa_{xx}(t)& =  \left[ \pa^2_x \Phi_0 \right]_{x_c(t)}\;, \\
		\kappa_{xy}(t)& =  \left[ \pa^2_{xy} \Phi_0\right]_{x_c(t)}\;, \\
		\kappa_{yy}(t)& =  \left[ \pa^2_y \Phi_0\right]_{x_c(t)}\;, \\
		F_x(t)& = -  \left[\pa_x \Phi_P\right]_{x_c(t)}\;, \\
		F_y(t)& = -  \left[\pa_y \Phi_P\right]_{x_c(t)} \;.
	\end{split}\end{equation}
In Eqs. \eqref{eq:sb2} it is assumed that $\lambda>0$ and that all the $\kappa$'s and $F$'s are given periodic functions of time with period $T$:
	\begin{equation}\begin{split}
	 	\kappa_{xx}(t)& = \kappa_{xx}(t+T)\;, \\
		\kappa_{xy}(t)& =  \kappa_{xy}(t+T)\;, \\
		\kappa_{yy}(t)& =   \kappa_{yy}(t+T)\;, \\
		F_x(t)& =  F_{x}(t+T)\;, \\
		F_y(t)& =  F_{y}(t+T) \;.
	\end{split}\end{equation}
The goal is to solve Eqs. \eqref{eq:sb2} to find $\bfx_1(t)=\left( x_1(t),y_1(t) \right)$. 
We want to find the solution of this equation that neglects transients, i.e., the solution to which 
all solutions tend in the limit $t\to\infty$. This solution is the one such that $\bfx_1(t)$ 
is periodic with the same periodicity of the $F$'s and $k$'s. 
This is also a requirement if we want to obtain a closed orbit,
and we have to assume such periodicity if our phenomenological model is to
make sense.\footnote{Note that Eqs. \eqref{eq:sb1} is conceptually similar to the following simpler equation
	\begin{equation}
		\ddot{x} + \kappa(t) x + \lambda \dot{x} = F(t)\;. \label{eq:osc1}
	\end{equation}
If $\kappa$ were constant and not time dependent, Eq.~\eqref{eq:osc1} would be the equation of a damped and
driven harmonic oscillator. The general solution of this equation is a
transient that decays exponentially plus a periodic term with the same
periodicity as $F(t)$. Our problem is more general, and $\kappa$ is a function of time. 
We are interested in the solution that neglects the transients, which is the generalisation of the solution of the damped and driven harmonic oscillator that neglects the decaying exponential. This solution is the one such that $x(t)$ is periodic with the same periodicity of
$F(t)$ and $k(t)$.} Let us therefore expand all time dependent quantities as Fourier series, assuming they are periodic with period $T$:
	\begin{equation}\begin{split}
	 	x_1(t) &= \sum_n X_n e^{i n \omega t}\;,  \\
		y_1(t) &= \sum_n Y_n e^{i n \omega t} \;, \\
		\kappa_{xx}(t) &= \sum_n K_{xx;n} e^{i n \omega t}\;,  \\
		\kappa_{xy}(t) &= \sum_n K_{xy;n} e^{i n \omega t}\;, \\
		\kappa_{yy}(t) &= \sum_n K_{yy;n} e^{i n \omega t} \;, \\  
		F_x(t) &=  \sum_n F_{x;n} e^{i n \omega t}\;, \\
		F_y(t) &=  \sum_n F_{y;n} e^{i n \omega t}\;. 
	\end{split}\end{equation} 
Fortunately, the product of the two Fourier series gives us another Fourier series of the same type. 
Substituting the Fourier expansions into Eqs. \eqref{eq:sb2} and then equating coefficients term by
term we arrive at
	\begin{equation}\begin{split}
		- \left[ (n \omega)^2 + \Omega_p^2 \right] X_n &=   \\
			 - \sum_{m} K_{xx;m} X_{n-m} &- \sum_{m} K_{xy;m} Y_{n-m}  - (i n \omega) \left[ 2 \lambda X_n + 2 \Omega_p Y_n \right]  + F_{x;n}\;, \\
		- \left[ (n \omega)^2 + \Omega_p^2 \right] Y_n &=   \\
			 - \sum_{m} K_{yy;m} Y_{n-m} &- \sum_{m} K_{xy;m} X_{n-m}  - (i n \omega) \left[ 2 \lambda Y_n - 2 \Omega_p X_n \right]  + F_{y;n} \;.
	\end{split}\end{equation}
The last two equations are a linear algebraic system of equations in the unknowns $X_n$ and $Y_n$. All the other Fourier coefficients, $K$'s and $F$'s, are assumed to be known. To solve this system let us rewrite it in the form 
\begin{equation} \mathbb{A} \mathbf{X} = \mathbf{F} \;, \label{eq:algsys}\end{equation}
where $\mathbb{A}$ is an infinite matrix and
\begin{equation} 
\mathbf{X}  = 
	 \begin{pmatrix}
	   \vdots \\ X_{-1} \\ Y_{-1} \\ X_0 \\  Y_0 \\ X_{1} \\ Y_{1} \\ \vdots 
	 \end{pmatrix}\;.
\end{equation}
$\mathbf{F}$ is
\begin{equation} 
\mathbf{F}  = 
	 \begin{pmatrix}
	   \vdots \\ F_{x;-1} \\ F_{y;-1} \\ F_{x;0} \\  F_{y;0} \\ F_{x;1} \\ F_{y;1} \\ \vdots 
	 \end{pmatrix}\;.
\end{equation}
To write $\mathbb{A}$, let us divide it into a part $\mathbb{K}$ containing the Fourier coefficients of the $\kappa's$ and a part $\mathbb{D}$ that does not:
  \begin{equation}
 \mathbb{A} = \mathbb{D} + \mathbb{K}\;.
 \end{equation}
 $\mathbb{D}$ is a block-diagonal matrix: it has
$2\times2$ blocks along the diagonal. Each block is 
\begin{equation}
\mathbb{D}_n = 
	 \begin{pmatrix}
		-(n \omega)^2 + 2 \lambda (i n \omega) + \Omega_p  & (in\omega) 2 \Omega_p \\
		- (in\omega) 2 \Omega_p & -(n \omega)^2 + 2 \lambda (i n \omega) + \Omega_p
	 \end{pmatrix}\;.
 \end{equation}
This is the block referring to the vector $(X_n,Y_n)$. The entire matrix $\mathbb{D}$ is then
\begin{equation}
\mathbb{D} = 
	 \begin{pmatrix}
	    \ddots  & &  & &  \\
	  &  \mathbb{D}_{-1}  & &  &  \\
	  & & \mathbb{D}_0  & & \\
	  & & &  \mathbb{D}_1  & \\
	  & & & &  \ddots  \\
	 \end{pmatrix}\;.
 \end{equation}
 The matrix $\mathbb{K}$ is 
 \begin{equation}
 \mathbb{K} =
	 \begin{pmatrix}
	 \ddots & \vdots & \vdots & \vdots & \vdots & \vdots & \ddots \\
	 \cdots & \mathbb{K}_{0} & \mathbb{K}_1 & \mathbb{K}_2 & \mathbb{K}_3 & \mathbb{K}_4 & \cdots \\
	 \cdots & \mathbb{K}_{-1} &  \mathbb{K}_{0} & \mathbb{K}_1 & \mathbb{K}_2 & \mathbb{K}_3 & \cdots \\
	 \cdots & \mathbb{K}_{-2} & \mathbb{K}_{-1} &  \mathbb{K}_{0} & \mathbb{K}_1 & \mathbb{K}_2 & \cdots \\
	 \cdots & \mathbb{K}_{-3} & \mathbb{K}_{-2} & \mathbb{K}_{-1} &  \mathbb{K}_{0} & \mathbb{K}_1 & \cdots \\
	 \cdots & \mathbb{K}_{-4} & \mathbb{K}_{-3} & \mathbb{K}_{-2} & \mathbb{K}_{-1} &  \mathbb{K}_{0} & \cdots \\
	 \ddots & \vdots & \vdots & \vdots & \vdots & \vdots & \ddots \\
	 \end{pmatrix}\;,
 \end{equation}
where each $\mathbb{K}_n$ is a $2\times2$ block given by
 \begin{equation}
\mathbb{K}_n =
	 \begin{pmatrix}
	  K_{xx;n} & K_{xy;n} \\
	  K_{xy;n} & K_{yy;n} 
	 \end{pmatrix}\;.
 \end{equation}
Since the linear algebraic system represented by Eq. \eqref{eq:algsys} is infinite, we need to truncate it in order to be able to solve it. On the diagonal of the matrix $\mathbb{K}$ there is $\mathbb{K}_0$. Suppose we truncate the Fourier expansion of all $\kappa$'s at order $\mathbb{K}_0$. Then the matrix $\mathbb{A}$ is block diagonal and the system is very easy to solve by considering only line pairs at a time. Now suppose we truncate the Fourier expansion of $\kappa(t)$ at some higher order. In this case, the matrix $\mathbf{A}$ is not block diagonal, but is a block band matrix that has elements on the sides of the diagonal up to the order of truncation. For example, if we truncate at $\mathbb{K}_{\pm 1}$, then we have a (block) band of width 3. To find the solution we need to truncate the system at some finite order and in general we cannot solve the system exactly as in the case in which we truncate at $\mathbb{K}_0$, where equations are decoupled in pairs. Thus there are two truncations involved. The first is where to truncate the Fourier expansions of $\kappa$ and $F$. This truncation has to be done at some order where $\kappa$ and $F$ are well represented by their Fourier expansions. The second is the size of truncation of the system $\mathbb{A}$, and must be done after the first truncation has been performed. This second truncation must be done for a size sufficiently high that the solution is converged and is not affected by a further increase in the size of the system. We have found that this convergence takes place quite rapidly. 

\end{document}